\documentclass[lettersize,journal]{IEEEtran}
\usepackage{amsmath,amsfonts}
\usepackage{algorithmic}
\usepackage{algorithm}
\usepackage{array}
\usepackage[caption=false,font=small,textfont=sf]{subfig}
\usepackage{textcomp}
\usepackage{stfloats}
\usepackage{url}
\usepackage{verbatim}
\usepackage{graphicx}
\usepackage{cite}
\usepackage{xcolor}
\usepackage{lipsum}
\usepackage{mathrsfs}

\newtheorem{theorem}{Theorem}
\newtheorem{problem}{Problem}

\newtheorem{definition}[theorem]{Definition}
\newtheorem{assumption}[theorem]{Assumption}

\newtheorem{remark}[theorem]{Remark}

\newtheorem{property}{$\mathfrak{P}$}

\begin{document}

\title{Uncovering the Influence Flow Model of Transistor Amplifiers, Its Reconstruction and Application}

\author{Mohammed Tuhin Rana,~\IEEEmembership{Graduate Student Member,~IEEE}, Mishfad Shaikh Veedu, Murti V. Salapaka,~\IEEEmembership{Fellow,~IEEE}}

\maketitle

\begin{abstract}
Multistage transistor amplifiers can be effectively modeled as network of dynamic systems where individual amplifier stages interact through couplings that are dynamic in nature. Using circuit analysis techniques, we show that a large class of transistor amplifiers can be modeled as Linear Dynamic Influence Model (LDIM), where the interactions between different amplifier stages are modeled as linear dynamic equations. LDIM modeling of transistor circuits leads to application of data-driven network reconstruction techniques to characterize stage interactions and identify faults and critical circuit parameters efficiently. Employing graphical modeling techniques and Wiener filtering, we demonstrate that the network structure can be reconstructed solely from voltage time-series measurements sampled at specified points in the circuit. The efficacy of these network reconstruction methods in multistage amplifiers is demonstrated through extensive simulations involving multiple amplifier circuits in Cadence, as well as experimental results on physical hardware.  The ability to infer network structure directly from measurement data offers designers and users efficient tools to design, analyze, and debug amplifier circuits. To demonstrate the utility of network reconstruction in multistage amplifier circuits, a fault diagnosis method leveraging these techniques is presented.
\end{abstract}


\section{Introduction}
\IEEEPARstart{T}{ransistor} amplifiers are prevalent in modern communication and signal processing devices (see e.g. \cite{razavi_rf_microelectronics,razaviCMOSIC}). Some of its applications include use in cellular communication \cite{Lee_cellular_application}, satellite communication \cite{Liufu_satellite_application}, medical devices \cite{Um_medical_application}, and consumer electronics \cite{Li_consumer_electronics_application}. Most such applications require interconnection of multiple amplifier stages. For instance, the authors in \cite{Zhang_satellite_application} use a cascode amplifier stage along with a differential pair to develop a low noise amplifier for satellite communications. An interconnection of three amplifier stages was used in \cite{Tombak_cellular_application} to develop a power amplifier for a cellular application. In many such applications the number of interconnected stages can be substantially high. For example, the authors of \cite{Jun-Chau_dist_amplifier} use nine stages in a $3\times 3$ distributed fashion. An eight-stage broadband amplifier was developed in \cite{THJang_8_stage} for use in millimeter-wave communications. As the number of stages grow in a multistage amplifier, their design and debugging process become more complex and challenging. In this article, we show that multistage electronic amplifiers can be interpreted as \emph{networks of dynamic systems} which can aid in design and debugging process of such amplifiers.
\par Network of dynamic systems have become an effective tool to describe complex modular systems. Some existing applications include biological sciences \cite{biology_del_modular}, cognitive science \cite{cognitive_quinn_estimating}, and economics \cite{economics_mantegna_introduction}. As described previously, most transistor amplifiers are built in a modular fashion where different stages are interconnected to achieve, for example, signal processing or amplification goal. Such modularity leads us to hypothesize that modeling and inference tools available in network of dynamic systems can be applicable in transistor amplifiers. Toward the verification of this hypothesis, we show that many transistor amplifiers can be modeled as a linear dynamic influence model (LDIM) \cite{materassi_Signal_selection}. In a preliminary study \cite{rana_causal_discovery_in_electronics}, we demonstrated through examples that the operating point equations of static BJT circuits can be interpreted to have structural properties which lead to a graphical model. When compared to \cite{rana_causal_discovery_in_electronics}, the method presented in this article can accommodate the considerably large class that include dynamic elements. Moreover, the improved modeling framework of this article is in terms of node voltages of the circuits rather than device operating points. Thus, making the data acquisition and reconstruction process much simpler and convenient compared to \cite{rana_causal_discovery_in_electronics}.

\par As a second contribution, we show that causal inference techniques used in network of dynamic systems can be applied to transistor amplifiers in order to infer the signal flow paths in such circuits. The applicability of such techniques in transistor amplifier circuits can open up new frontiers for design and debugging methodologies. Moreover, we show that the reconstruction of network structure from data of such amplifier systems can help in fault diagnosis.

\par Fault diagnosis of analog electronic circuits is a challenging task \cite{BINU_fault_diag_review}. According to \cite{BINU_fault_diag_review} most fault diagnosis algorithms can be categorized into two categories, namely: simulation before test (SBT) and simulation after test (SAT). The SBT methods simulate an extensive number of possible fault scenarios to build a library or statistical database which is then used in different fault diagnosis methods. Such methods include verifications against mathematical model \cite{Model_Based_SBT}, meta heuristic algorithm based methods \cite{Optimization_PSO_SBT}, and machine learning based methods \cite{Machinelearning_NN_SBT}. In case of SBT the fault diagnosis task becomes a search problem after building the fault library, but the method requires an extensive data set which leads to requirement of large amount of memory and computational resources. On the other hand, as the name suggests, the SAT methods perform simulation of the circuit after test data is obtained \cite{SVM_SAT}. These methods require less resources compared to the SBT methods. However, simulation of different fault scenarios still require significant amount of computational resources. In this article we show that network reconstruction based approach can be used for fault diagnosis of analog circuits directly from measurement data drastically reducing memory and computational capacity requirement.

\par Apart from its use in circuit design and debugging, the findings of this article lay the foundations for the development of a benchmark platform for testing and evaluating \emph{causal inference} algorithms. The research on causal inference strives to uncover causal relationships among measured quantities of networked systems \cite{Peter_clark_book}. However, it is challenging to evaluate such algorithms in real-world data due to the scarcity of real-world datasets where the ground truth about the causal relationships are known. Which is why real-world datasets with known ground truths are in high demand \cite{runge_causeme_1,runge_causeme_2}. The modeling of transistor amplifiers as LDIMs leads to the uncovering of the ground truth of the underlying influence flow structure of such circuits. Thus, transistor amplifiers are prime candidates for developing benchmark platforms for evaluating causal inference algorithms.  

\par The key contributions of this article are as follows: $1)$ We show that a large class of amplifier networks can be modeled as LDIM. These amplifiers include common source, source follower, and cascode amplifiers which are widely used in different applications (see also \cite{THJang_8_stage,li3STIA,Chen_Cascode,Fujishima_common_source}). $2)$ We demonstrate the reconstruction of the underlying signal-flow structure entailed by the LDIM of the transistor amplifier. $3)$ A simple procedure for fault diagnosis based on the reconstruction method is demonstrated and other potential applications areas are noted.

\par The article is organized as follows: Section \ref{sec: preliminaries} presents preliminaries that aid the discussions in the rest of the article. Section \ref{sec: LDIM of Transistor Amplifiers} shows that a large class of transistor amplifiers can be modeled as network of dynamic systems. Section \ref{sec: Simulation Results} and Section \ref{sec: Experimental Results} present simulation and experimental results of network reconstruction in transistor amplifiers establishing the efficacy of the modeling and reconstruction approach for such circuits. Section \ref{sec: Application Note} shows an application where we show that the network reconstruction of amplifier circuits from data can aid in the fault diagnosis process. Finally, Section \ref{sec: Conclusion} concludes the article.

\section{Preliminaries}\label{sec: preliminaries}

\par In this section we recall preliminary material on circuit theory, graph theory, and causal discovery. There is some overlap in terminologies employed in circuit theory and graph theory. For example, the notions \emph{node} and \emph{path} are used in both of these disciplines with different meanings. We use the following definitions to delineate the differences.

\subsection{Notions of Graph Theory and Causal Inference:}
\par A few notions of graph theory and causal inference are recalled from \cite{materassi_Wiener_filter}, and \cite{J_Pearl_Causal_inf_book} in this section. First, basic definitions of graph theory are presented and then notions of causal inference and networked systems are recalled.
\begin{definition}\emph{(Directed, Undirected, and Partially Directed Graph):}
    A graph $\vec{\mathcal{G}}:=(\mathcal{V},\vec{\mathcal{E}})$, where $V$ is a set of vertices and $\vec{\mathcal{E}}$ is the set of edges, is called a directed graph if each element of $\vec{\mathcal{E}}$ is an ordered pair $(x,y)\in \mathcal{V}\times \mathcal{V}$.  Similarly for a graph $\mathcal{G}:=(\mathcal{V},\mathcal{E})$ if all the elements of $\mathcal{E}$ are unordered pair of elements from $\mathcal{V}$ then $\mathcal{G}$ is called an undirected graph. Note that, we denote the directed edge $x\rightarrow y$ by $(x,y)$, and the undirected edge $x-y$ by $\{x,y\}$. If some of the edges of a graph is oriented and others are not oriented then the graph is called a partially directed graph.
\end{definition}
\begin{definition}[Skeleton]
    The skeleton of a directed graph $\vec{\mathcal{G}}=(\mathcal{V},\vec{\mathcal{E}})$, is obtained by removing the orientation of the edges in $\vec{\mathcal{E}}$; the skeleton is denoted by ${\mathcal{G}_S=(\mathcal{V},\mathcal{E}_S)}$.
\end{definition}
\begin{definition}[Chain]
    Given a directed graph $\vec{\mathcal{G}}=(\mathcal{V},\vec{\mathcal{E}})$, a chain from vertex ${x}$ to ${y}$ is a sequence of vertices ${\{\pi_{k}\}_{k=1}^{n}}$, such that, ${(\pi_{l},\pi_{l+\mathrm{1}})\in \vec{\mathcal{E}}}$ ${\forall \mathrm{1}\leq l\leq n}$, where, ${\pi_{\mathrm{1}}=x}$ and ${\pi_{n}=y}$.
\end{definition}
\begin{definition}[Cycle]
    In a directed graph $\vec{\mathcal{G}}$, a cycle is a chain from a vertex ${x}$ to ${x}$ itself.
\end{definition}
\begin{definition}[Path]
    Given a directed graph $\vec{\mathcal{G}}=(\mathcal{V},\vec{\mathcal{E}})$, a path from vertex ${x}$ to ${y}$ is a sequence of vertices ${\{\pi_{k}\}_{k=1}^{n}}$, such that, either ${(\pi_{l},\pi_{l+\mathrm{1}})\in \vec{\mathcal{E}}}$ or ${(\pi_{l+\mathrm{1}},\pi_{l})\in \vec{\mathcal{E}}}$ holds for all ${\mathrm{1}\leq l\leq n}$,  where, ${\pi_{\mathrm{1}}=x}$ and ${\pi_{n}=y}$.
\end{definition}
\begin{definition}[Directed Acyclic Graph]
    A directed graph $\vec{\mathcal{G}}$ with finite number of vertices and no cycles is called a directed acyclic graph (DAG).
\end{definition}
\begin{definition}[Parent and Child]
    Given a directed graph $\vec{\mathcal{G}}=(\mathcal{V},\vec{\mathcal{E}})$, vertex ${x}$ is called a parent of ${y}$ if ${(x,y)\in \vec{\mathcal{E}}}$, and in that case ${y}$ is called a child of ${x}$.
\end{definition}
\begin{definition}[Ancestor and Descendant]
     In a directed graph $\vec{\mathcal{G}}$, vertex ${x}$ is called an ancestor of ${y}$ if there is a directed chain from ${x}$ to ${y}$ or $x=y$, and in that case ${y}$ is a descendant of ${x}$.
\end{definition}
\begin{definition}[Adjacent vertices]
    In a graph ${\mathcal{G}=(\mathcal{V},\mathcal{E})}$, two vertices ${x}$ and ${y}$ are said to be adjacent, if one of the following conditions satisfy, $1)$${(x,y) \in \mathcal{E}}$, $2)$${(y,x) \in \mathcal{E}}$, $3)$${\{x,y\} \in \mathcal{E}}$. The notation ${adj(x,y)}$ denotes ${x}$ and ${y}$ are adjacent, and ${\sim adj(x,y)}$ denotes $\mathnormal{x}$ and $\mathnormal{y}$ are not adjacent.
\end{definition}
\begin{definition}[Adjacency]
    \emph{Adjacency}${(\mathcal{G},x)}$ is defined as the the set of vertices adjacent to ${x}$ in the graph $\mathcal{G}$.
\end{definition}
\begin{definition}[Collider and Fork]
    Consider a directed graph $\vec{\mathcal{G}}=(\mathcal{V},\vec{\mathcal{E}})$, a path involving vertices $(\pi_{{1}}...\pi_{m})$ has a collider at ${\pi_{l}}$ if ${(\pi_{l-\mathrm{1}},\pi_{l})\in \vec{\mathcal{E}}}$ and ${(\pi_{l+\mathrm{1}},\pi_{l})\in \vec{\mathcal{E}}}$ holds (i.e. $\pi_{l-1}\rightarrow\pi_l\leftarrow\pi_{l+1}$ is in the path). And a path has a fork at ${\pi_{l}}$ if ${(\pi_{l},\pi_{l-\mathrm{1}})\in \vec{\mathcal{E}}}$ and ${(\pi_{l},\pi_{l+\mathrm{1}})\in \vec{\mathcal{E}}}$ holds (i.e. $\pi_{l-1}\leftarrow\pi_l\rightarrow\pi_{l+1}$ is in the path).
\end{definition}
\begin{definition}[v-structure]
    A collider whose parents do not have an edge between them is called a \emph{v-structure}.
\end{definition}
\begin{definition}[d-Separation]
    Given a directed graph $\vec{\mathcal{G}}:=(\mathcal{V},\vec{\mathcal{E}})$, two vertices ${x}$ and ${y}$ are said to be d-separated by a set ${Z\subset \mathcal{V}}$ if at least one of the following holds for all paths between $x$ and $y$.
    \begin{enumerate}
        \item The path contains a vertex ${p\in Z}$ which is not a collider.
        \item If the path contains a collider at a vertex ${q}$, then neither ${q}$ nor any of its descendants belong to the set ${Z}$.
    \end{enumerate}
    \par The notation $dsep{(x,Z,y)}$ indicates that ${x}$ and ${y}$ are d-separated by ${Z}$. The sets ${X\subset \mathcal{V}}$, ${Y\subset \mathcal{V}}$, ${Z\subset \mathcal{V}}$ are said to be d-separated if ${dsep(x,Z,y)}$ holds for every ${x\in X}$ and ${y\in Y}$.
\end{definition}

\par Next we discuss graphical models of networked systems and its reconstruction from observational data. To that end one requires a \emph{d-separation oracle} that can answer questions on d-separation from data. In static networks conditional independence test such as \emph{Fisher-Z test} \cite{kalisch_fisher_z} can be used as a \emph{d-separation oracle}. However, such tests are inadequate in dynamic networks. In linear dynamic networks such a \emph{d-separation oracle} can be constructed using properties of \emph{Wiener Filters} \cite{materassi_Signal_selection}. Below is a brief description on Wiener filters and its use in network reconstruction.
\subsection{Network of Dynamic Systems and Wiener Filter:}

\par Here we recall notions of network of dynamic systems from \cite{materassi_Wiener_filter,materassi_Signal_selection}.

\begin{definition}\emph{(Wide-sense Stationary (WSS) and Jointly WSS Process):} Consider a vector of time-discrete stochastic process $x(\cdot)$. $x(\cdot)$ is a (WSS) process if for all $t_1,t_2\in\mathbb{Z}$ such that $t_2=t_1+\delta$, $\mathbf{E}[x(t_1)]=\mathbf{E}[x(t_2)]$ and $\mathbf{E}[x(t_1)x^T(t_2)]=\mathbf{E}[x(0)x^T(\delta)]$ where, $\mathbf{E[.]}$ is the expectation operator. Two WSS vector processes $x(\cdot)$ and $y(\cdot)$ are jointly WSS if $\mathbf{E}[x(t_1)y^T(t_2)]=\mathbf{E}[x(0)y^T(\delta)]~\forall t_1, t_2$.
\end{definition}

\begin{definition}[Power Spectral Density]
    The power spectral density (PSD) of a WSS vector process $x(\cdot)$ is defined as
    \begin{align}
        &\Phi_{xx}(\omega):=\sum_{\delta=-\infty}^{\infty}\mathbf{E}[x(0)x^T(\delta)]e^{-\hat{j}\omega\delta}.
    \end{align}
    The cross power spectral density (CPSD) of time-discrete jointly WSS vector processes $x(\cdot)$ and $y(\cdot)$ is defined as
    \begin{align}
        &\Phi_{xy}(\omega):=\sum_{\delta=-\infty}^{\infty}\mathbf{E}[x(0)y^T(\delta)]e^{-\hat{j}\omega\delta}.
    \end{align}
\end{definition}

\par Next we recall a result from \cite{materassi_Wiener_filter} and \cite{materassi_Signal_selection} on Wiener filtering for jointly WSS process.

\begin{theorem}
    Let the $x(\cdot)$ and $y(\cdot)=[y_1(\cdot)~y_2(\cdot)...~y_n(\cdot)]^T$ be scaler and $n$ dimensional vector processes respectively that are jointly WSS. Let  and $\chi:=\sum_{k=1}^{n}\sum_{t=-\infty}^{\infty}c_ty_k(t)$, where $c_t\in\mathbb{R}$ for all $t$. Consider the problem of estimating $x(\cdot)$ from $y(\cdot)$:
    \begin{align}
        x_y(t)=\arg\inf_{\substack{q\in \chi}}\mathbf{E}\left[(x(t)-q)^2\right],~~t\in\mathbb{Z}.\nonumber
    \end{align}
     If the PSD matrix $\Phi_{yy}(\omega)$ is positive definite for $\omega\in[-\pi,\pi]$, then the solution to the problem exist, is unique and is given in frequency domain by $\hat{X}_y(\omega)=W_{x|y}(\omega)\hat{Y}(\omega)$, where $\hat{X}(\omega)$ denotes the Fourier transform of the signal $x(\cdot)$. $W_{x|y}(\omega)$ (the Wiener filter) is given by:
     \begin{align}
         W_{x|y}(\omega) = \Phi_{xy}(\omega)\Phi_{yy}^{-1}(\omega).\nonumber
     \end{align}
\end{theorem}

\begin{definition}
    Let $x(\cdot)$ and $y(\cdot)=[y_1(\cdot)~y_2(\cdot)...~y_n(\cdot)]^T$ be scaler and $n$ dimensional processes respectively, that are jointly WSS. Let $W_{x|y}(\omega)$ be the Wiener filter for estimating $x(\cdot)$ from $y(\cdot)$ with $\hat{X}_y(\omega)=W_{x|y}(\omega)\hat{Y}(\omega)=\left[W_{x,[y_1]|y}(\omega)~...~W_{x,[y_n]|y}(\omega)\right][\hat{Y}_1(\omega)~...~\hat{Y}_n(\omega)]^T$.
\end{definition}

\begin{definition}[Wiener Separation \cite{materassi_Signal_selection}]\label{def: wsep}
    Let $x(\cdot),~ y(\cdot)$ be two scaler random process and $Z$ be a set of random process that are jointly WSS such that $x,y\notin Z$. We say that $x(\cdot)$ is Wiener-separated from $y(\cdot)$ by $Z$ if $W_{y,[x]|(x, Z)}(\omega)=0$ for all $\omega\in[-\pi,\pi]$, and in that case we say $wsep(x,Z,y)$ holds. Note that $W_{y,[x]|(x, Z)}(\omega)$ denotes the component of the Wiener filter operating on $x(\cdot)$ when estimating $y(\cdot)$ from the processes in $\{x(\cdot)\}\cup Z$.
\end{definition}

\par It was shown in \cite{materassi_reconstruction_LDDAG} that Wiener-separation is symmetric i.e. $wsep(x,Z,y)\iff wsep(y,Z,x)$.

\begin{definition}\label{def: LDIM Definition}\emph{(Linear Dynamic Influence Model (LDIM)) \cite{materassi_Signal_selection}:} Consider a network of dynamic system that can be expressed as
    \begin{align}
        \hat{Y}(\omega)=\hat{\mathcal{H}}(\omega)\hat{Y}(\omega)+\hat{\varepsilon}(\omega),\label{eq: LDIM definition}
    \end{align}
    where $\hat{Y}(\omega)$ is Fourier transform of a $n\times1$ vector of time-series measurements, $\hat{\mathcal{H}}(\omega)$ is a $n\times n$ stable transfer matrix, and $\hat{\varepsilon}(\omega)$ is Fourier transform of a $n\times1$ vector of wide-sense stationary inputs with diagonal $\Phi_{{\varepsilon}{\varepsilon}}(\omega)$. Such a network is called an LDIM and is denoted by $(\hat{\mathcal{H}},\hat{\varepsilon})$. 
\end{definition}

\begin{definition}[Generative Graph associated with LDIM]
    Consider an LDIM $(\hat{\mathcal{H}},\hat{\varepsilon})$ with output process $y(t)=[y_1(t)~y_2(t)~...~y_n(t)]^T$ and a directed graph $\vec{\mathcal{G}}=(\mathcal{V},\vec{\mathcal{E}})$ with $n$ vertices. With slight abuse of notation we denote the vertices of $\vec{\mathcal{G}}$ as $\mathcal{V}=\{y_1,~y_2,~...~y_n\}$. $\vec{\mathcal{G}}$ is called generative graph associated with the LDIM $(\hat{\mathcal{H}},\hat{\varepsilon})$ if $\vec{\mathcal{E}}=\{(y_i,y_j)| \hat{\mathcal{H}}_{ji}(\omega)\neq0\}$. We often refer to the generative graph as the generative structure.
\end{definition}

\par The following result from \cite{materassi_Signal_selection} establishes a connection between d-separation in generative graphs and Wiener separation associated with LDIM.

\begin{theorem}
    Consider an LDIM with graphical representation $\vec{\mathcal{G}}=(\mathcal{V},\vec{\mathcal{E}})$. Let $x,y\in \mathcal{V}$ and $Z\subseteq \mathcal{V}\setminus\{x,y\}$, then $dsep(x,Z,y)\implies wsep(x,Z,y)$.
\end{theorem}

\par In addition to the above result the assumption of faithfulness ensures the converse relationship is also true.

\begin{assumption}
    Consider an LDIM with graphical representation $\vec{\mathcal{G}}=(\mathcal{V},\vec{\mathcal{E}})$. Let $x,y\in \mathcal{V}$ and $Z\subseteq \mathcal{V}\setminus\{x,y\}$, then $dsep(x,Z,y)\impliedby wsep(x,Z,y)$.
\end{assumption}

\par Note that the Wiener filters can be obtained from measurement data using the estimated power spectral densities. They can also be directly estimated by using projection-based approach as shown in \cite{veedu_wiener_pc}. Thus under the faithfulness assumption the Wiener separation test can be used as a \emph{d-separation oracle} for causal discovery algorithms in LDIMs.

\par Next we discuss the Peter-Clarke (PC) algorithm from \cite{Peter_clark_book}. The PC algorithm is used in this article to reconstruct the underlying generative graph of electronic circuits from measurement data. A pseudo code of the PC algorithm is shown in Algorithm \ref{alg: PC_Pseudocode}. The PC algorithm takes time-series data as input and uses a \emph{d-separation oracle} to reconstruct the skeleton, v-structures, and some additional edge directions in three distinct steps as follows:

\begin{enumerate}
    \item reconstruct the skeleton of underlying DAG by using pairwise separation test,

    \item identify all the v-structures,

    \item if possible orient other edges by checking if a particular orientation of an edge creates a new v-structure.
\end{enumerate}

\par In this article we augment the PC algorithm with Wiener separation test as the d-separation oracle as shown in \cite{veedu_wiener_pc}. We estimate the Wiener filters using methods shown in \cite{veedu_wiener_pc}. To deal with finite samples error in the Wiener filter estimates we average the Wiener filters over a frequency range $\Psi=\{\omega_r,...,\omega_s\}$. The average magnitudes were then compared with a chosen threshold $\rho$ to determine the results of the Wiener separation test i.e. $\frac{1}{|\Psi|}\sum_{\omega_k\in\Psi}|W_{y,x|(x,Z)}(\omega_k)|<\rho\implies wsep(x,Z,y)$.
   \begin{algorithm}
      \caption{Peter-Clark Algorithm}\label{alg: PC_Pseudocode}
      \begin{algorithmic}
        \STATE Form a complete undirected graph ${\mathcal{G}=(\mathcal{V},\mathcal{E})}$.
        \STATE Set ${\eta={0}}$.
        \REPEAT
            \REPEAT
                \STATE Select an ordered pair of \emph{adjacent} vertices ${x,y\in \mathcal{V}}$, such that, cardinality of $|$\emph{Adjacency}${(\mathcal{G},x)\setminus \{y\}}|$ $\geq$ ${\eta}$.
                \REPEAT
                    \STATE Select $\mathnormal{Z\subseteq}$ \emph{Adjacency}${(\mathcal{G},x)\mspace{-3mu}\setminus \mspace{-3mu}\{y\}}$ with cardinality ${\eta}$.
                    \IF{$wsep(x,Z,y)$ holds}
                        \STATE Remove the edge between ${x}$ and ${y}$ in $\mathcal{G}$.
                        \STATE Save ${Z}$ as ${Z_{xy}}$.
                        \STATE break.
                    \ENDIF
                \UNTIL{all $\mathnormal{Z\subseteq}$ \emph{Adjacency}${(\mathcal{G},x)\setminus \{y\}}$ with cardinality ${\eta}$ has been considered.}
            \UNTIL{all adjacent pairs of vertices ${x}$ and ${y}$ with cardinality of \emph{Adjacency}${(\mathcal{G},x)\setminus \{y\}}$ $\geq$ ${\eta}$ has been considered.}
            \STATE ${\eta=\eta+{1}}$.
        \UNTIL{all adjacent pairs of vertices ${x}$ and $\mathnormal{y}$ have cardinality of \emph{Adjacency}${(\mathcal{G},x)\setminus \{y\}}$ $\le$ ${\eta}$.}        
        \REPEAT
            \STATE Choose ${a,b,c\in \mathcal{V}}$, such that, ${adj(a,c)}$, ${adj(b,c)}$, but ${\sim adj(a,b)}$ in ${\mathcal{G}}$.
            \IF{${c\notin Z_{ab}}$}
                \STATE Designate ${c}$ as a collier in the path from ${a}$ to ${b}$ in ${G}$.
            \ENDIF
        \UNTIL{all triplets ${a,b,c\in \mathcal{V}}$, such that, ${adj(a,c)}$, ${adj(b,c)}$, but ${\sim adj(a,b)}$ in $\mathcal{G}$ has been considered}
        \REPEAT
            \IF{${(a,b)\in \mathcal{E}}$, and ${adj(b,c)}$, and ${\sim adj(a,c)}$, and ${b}$ is not a child of another vertex}
                \STATE Designate ${b}$ as a parent of ${c}$
            \ENDIF
            \IF{there is a directed path from ${a}$ to ${b}$, and ${\{a,b\}\in \mathcal{E}}$}
                \STATE Designate ${a}$ as a parent of ${b}$
            \ENDIF
        \UNTIL{no more edges could be oriented.}
        
      \end{algorithmic} 
   \end{algorithm}
\par Next we present preliminary definitions from circuit theory which are summarized from \cite{thomasCircuitanalysis}. Even though, the reader might be thoroughly acquainted with these definitions, we reiterate them here to present the proposed ideas in the article in a self-contained manner and thus avoid ambiguity.

\subsection{Notions of Circuit Theory \cite{thomasCircuitanalysis}:}
\par In this subsection, we recall preliminary notions that are functional for later discussions in the article.

\begin{definition}[Circuit Elements]
    Circuit elements are models of electrical components. The circuit elements considered in this article are resistors, inductors, voltage sources, current sources, and transistors.
\end{definition}

\begin{definition}[Electrical Nodes]
    An electrical node in the model of a circuit is a point where more than one circuit element is connected together.
\end{definition}

\begin{definition}[Ground Node]
    The ground node of a circuit is the reference electrical node which is considered to be at zero potential.
\end{definition}

\begin{definition}[Power Node]
    An electrical node in a circuit where a power supply is connected.
\end{definition}

\begin{definition}[Electrical Path]
    An electrical path is a sequence of electrical nodes connected by circuit elements.
\end{definition}

\begin{definition}[Direct RLC Path]
    An electrical path consisting of only resistors, inductors, and capacitors is referred to as a direct RLC path if it does not contain ground or power nodes as one of the intermediate nodes.
\end{definition}
\section{Influence Flow Model of\\Transistor Amplifiers}\label{sec: LDIM of Transistor Amplifiers}
\par In this section we show that under certain assumptions transistor amplifiers can be modeled as a network of dynamic systems, where the agents influence each other through dynamic transfer function links. To maintain brevity we only discuss the results for MOSFET amplifiers, however, similar results hold for BJTs as well. 
\subsection{MOSFET Small Signal Noise Equivalent Model}
\par
\begin{figure}[!t]
    \centering
    \subfloat[]{\includegraphics[scale=0.7]{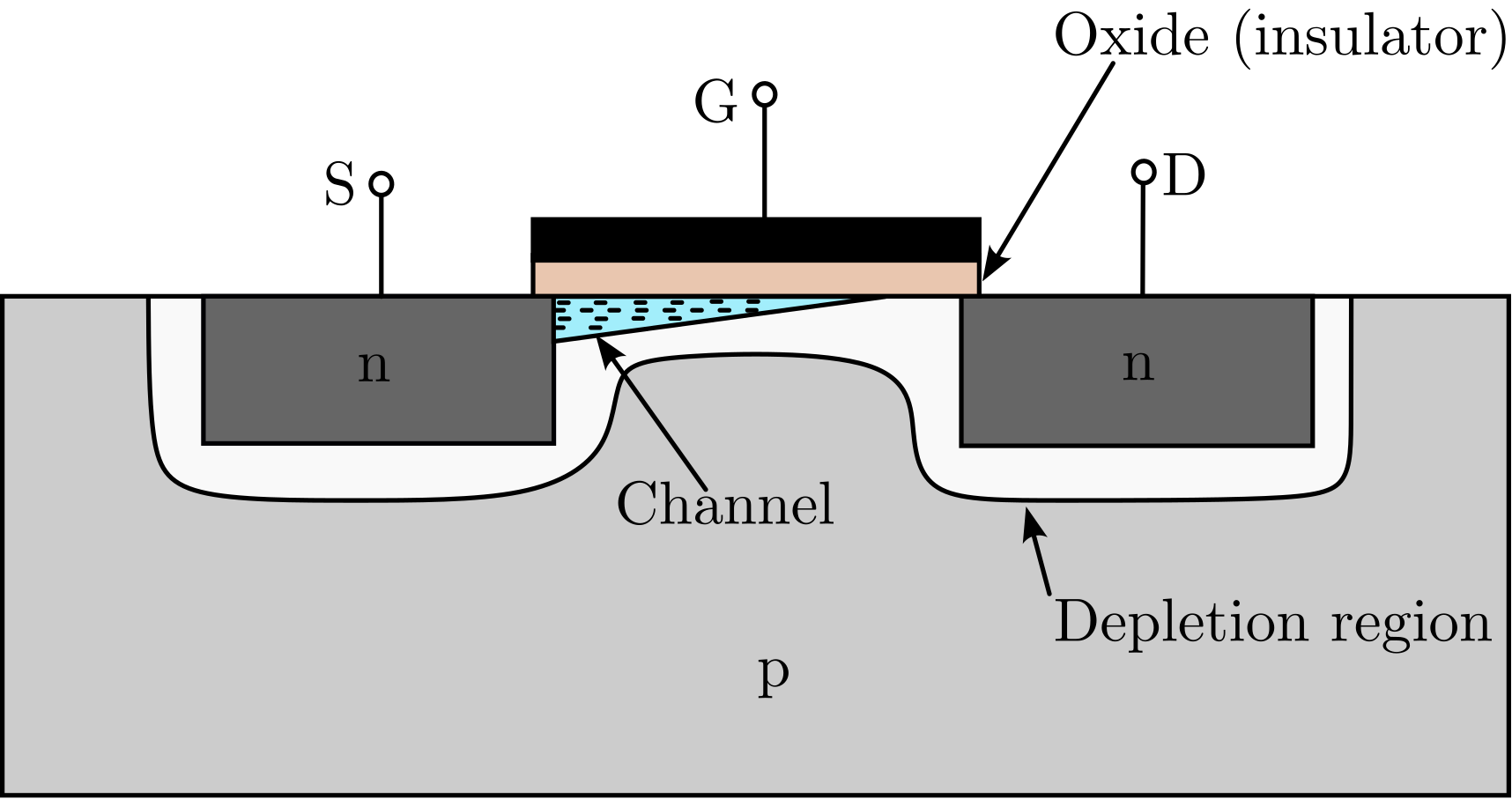}\label{fig: MOSFET 2D Layout}}
    \hfill
    \subfloat[]{\includegraphics[scale=1]{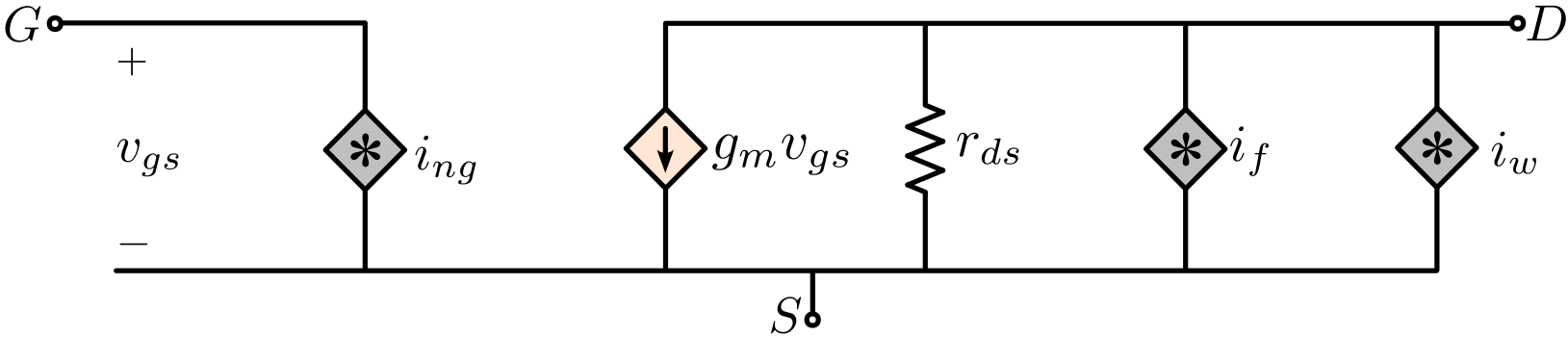}\label{fig: MOSFET Noise Equivalent}}
    \hfill
    \caption{(a) 2 dimensional structure of MOSFET. (b) Small signal noise equivalent model of MOSFET.}
    \label{fig: MOSFET 2D and Noise Equivalent}
\end{figure}

 We start by describing small signal noise equivalent model of a MOSFET. Fig. \ref{fig: MOSFET 2D Layout} shows the layout of an n-channel MOSFET; its small signal noise equivalent model is shown in Fig. \ref{fig: MOSFET Noise Equivalent} (see also \cite{motchenbacher1993low}). In the small signal model, $g_{m}$ represents the transconductance of the device, and $r_{ds}$ is due to channel length modulation of MOSFET (see also \cite{razaviCMOSIC}). Since other parasitic elements of the device are small and negligible at low or medium frequencies as shown in \cite{razaviCMOSIC}, we ignore their effect in this discussion. The small signal noise equivalent model of MOSFET contain three separate noise sources to represent the noise processes of the device. They are as (i) the channel between the source and the drain is resistive in nature and hence gives rise to thermal noise between them. The thermal noise is represented by the current source model $i_{w}$; (ii) The impurities at the channel surface lead to flicker noise which is represented by $i_{f}$ in the model; (iii) The noise of the gate insulator is represented by $i_{ng}$.
\subsection{MOSFET Biasing}
\begin{figure}[!t]
    \centering
    \includegraphics[scale=1]{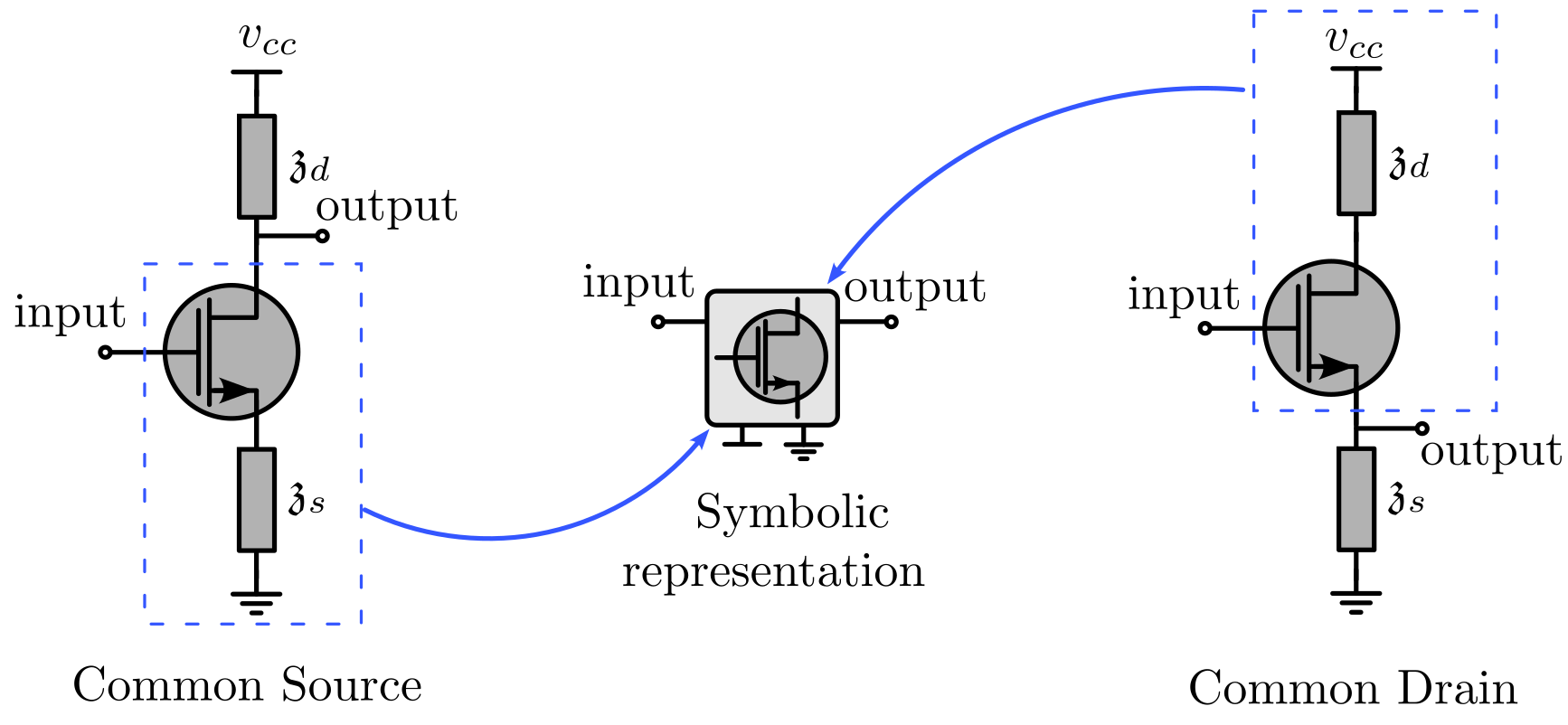}
    \caption{MOSFET in common source and common drain mode and its representative symbol.}
    \label{fig: Amplifier stage}
\end{figure}
\par It is crucial to establish the operating point of MOSFET amplifiers by means of proper biasing (see also \cite{boylestadElectronic}). We assume that MOSFETs are biased properly so that they operate in saturation region which is done in typical amplifier applications. Additionally, we consider two specific configurations, namely common source (CS) and source follower, also called common drain (CD), without any feedback bias as shown in Fig. \ref{fig: Amplifier stage}. In either case, we represent the portion inside the box with the symbol shown in Fig. \ref{fig: Amplifier stage} and call it an \emph{amplifier stage}. In the sequel, the input and output terminals of the amplifier stage is referred to as \emph{input port} and \emph{output port} respectively. Since an amplifier stage contains only one transistor we use the terms amplifier stage and transistor stage interchangeably. An electrical node where an output port of an amplifier stage is connected is referred to as an \emph{output node} of the circuit.
\par In addition to the amplifier stage, practical amplifiers contain passive subcircuits that serve various purpose such as, biasing, filtering, coupling between stages, DC blocking, and impedance matching (see also \cite{boylestadElectronic,razaviCMOSIC,razavi_rf_microelectronics}). We refer to these passive subcircuits as \emph{RLC blocks}. Such subcircuits are used to provide signal paths through direct RLC paths between output and input of consecutive stages, or between signal source and input of an amplifier stage. 
\par Utilizing the notions described above, it can be inferred that the amplifier circuits, where all the individual amplifier stages are in CS or CD configuration, can be interpreted to be an interconnection of RLC blocks and transistor stages. With the aforementioned structural interpretation of the circuits under consideration, we present the assumptions that are necessary for the analysis presented in this article.
\subsection{Assumptions}\label{sec: Assumptions}
The following assumptions on the structure of the circuit are made.
\begin{assumption}\label{assump: no rlc between outputs}
    There is no direct RLC path between two transistors' outputs ports.
\end{assumption}
\begin{remark}
    Direct connections with negligible parasitic resistance, inductance, capacitance are not considered as RLC elements in this article.
\end{remark}
\begin{assumption}\label{assump: no rlc between same transistor terminals}
    There is no direct RLC path between the gate and drain or gate and source of the same transistor.
\end{assumption}
\begin{remark}
    Assumption \ref{assump: no rlc between outputs} and Assumption \ref{assump: no rlc between same transistor terminals} ensure unilateral flow of signals which leads to a directed graphical model as shown in later sections.
\end{remark}
\begin{remark}
    Many practical circuits such as multistage radio frequency amplifiers, tuned amplifiers, and trans-impedance amplifiers satisfy these assumptions (see e.g. \cite{motchenbacher1993low,THjang8Stage,hasanCMOSbandpass}).
\end{remark}
\par The following assumptions are necessary for the analysis of the circuits to obtain the influence flow model.
\begin{assumption}\label{assump: infinite input impedance}
    The transistor amplifier stages have high input impedance.
\end{assumption}
\begin{assumption}\label{assump: small gate noise}
    The gate noise of the transistors are negligible.
\end{assumption}
\begin{assumption}\label{assump: noiseless passives}
    The passive components are either noiseless or produce negligible noise.
\end{assumption}
\begin{remark}
    As described in \cite{razaviCMOSIC}, MOSFET amplifiers in CS, CD, and Cascode configurations have high input impedance which validates Assumption \ref{assump: infinite input impedance}. This is a common consideration in amplifier design which is employed to make sure that amplifiers do not load the signal source, as such loading can lead to compromise in the signal integrity.
    Also, it is well known that gate noise is much smaller compared to the other noise in MOSFETs (see also \cite{grayAnalogICDesign}). Thus validating Assumption \ref{assump: small gate noise}.
    Furthermore, capacitors and inductors including resistors used in electronic circuits have an equivalent noise current that is negligible compared to that of the transistors' channel thermal noise and flicker noise. Hence, Assumption \ref{assump: noiseless passives} holds.
\end{remark}
\begin{remark}
    Its is also to be noted that Assumptions \ref{assump: infinite input impedance} - \ref{assump: noiseless passives} are used to simplify the analysis and modeling of the circuits. In the simulations and experimental results these assumptions are relaxed.
\end{remark}
In addition to the above we also need the following assumption to present the solution to the problem presented in this article.
\begin{assumption}\label{assump: Independent noise process}
    The noise process of the transistors are independent.
\end{assumption}
\begin{remark}
    In transistors, the processes that give rise to thermal and flicker noise are independent. Hence, the thermal and flicker noise of the transistors are independent \cite{grayAnalogICDesign}.
\end{remark}
\par With the preliminaries and the assumptions described, we next present the problem statement for modeling and reconstruction of transistor amplifier circuits.
\subsection{Problem Statement}
\par In this article we are concerned with two main problems; modeling transistor amplifier networks as linear dynamic influence models (LDIM), and its reconstruction using time-series measurements taken from the circuit. A formal description of the problem statements are given below: 
\begin{problem}\label{prob: LDIM Model}
    Given a amplifier circuit that satisfies the assumptions in Sec. \ref{sec: Assumptions} and has $n$ distinct output nodes, express the voltage of the output nodes of the circuit as an LDIM:
    \begin{align}\label{eq: voltage LDIM}
        \hat{V}(\omega)=\hat{\mathcal{H}}(\omega)\hat{V}(\omega)+\hat{\varepsilon}(\omega),
    \end{align}
    where $\hat{V}(\omega)$ is a $n\times1$ vector of Fourier transform of the sampled voltages of the output nodes, $\hat{\mathcal{H}}(\omega)$ is a $n\times n$ transfer matrix, and $\hat{\varepsilon}(\omega)$ is a $n\times1$ vector of noise processes of the transistors such that its cross power spectral density matrix is diagonal.
\end{problem}
\begin{problem}\label{prob: reconstructiom}
    Reconstruct the underlying generative graph associated with the LDIM model of an electronic circuit using voltage measurements at the output nodes of the circuit. 
\end{problem}
\par We utilize popular tools from graphical modeling and network reconstruction to solve Problem \ref{prob: reconstructiom} (see \cite{materassi_Wiener_filter,veedu_wiener_pc}). We then show that such reconstruction can aid in fault diagnosis.
\par To streamline the solution of Problem \ref{prob: LDIM Model}, we discuss structural properties of the amplifier circuits that satisfy Assumptions \ref{assump: no rlc between outputs} and \ref{assump: no rlc between same transistor terminals}.
\subsection{Structural Properties of Considered Circuits}
\par Assumptions \ref{assump: no rlc between outputs} and \ref{assump: no rlc between same transistor terminals} along with general considerations in amplifier design, lead to structural properties of the considered circuits as discussed below.
\begin{property}
    An RLC block can be connected to at most one output node.
\end{property}
 \par If more than one output nodes are connected to a RLC block then there would be a direct RLC path between them and hence between output ports of two different transistors, which is a violation of Assumption \ref{assump: no rlc between outputs}.

\par
\begin{figure}[!t]
    \centering
    \subfloat[]{\includegraphics[scale=1]{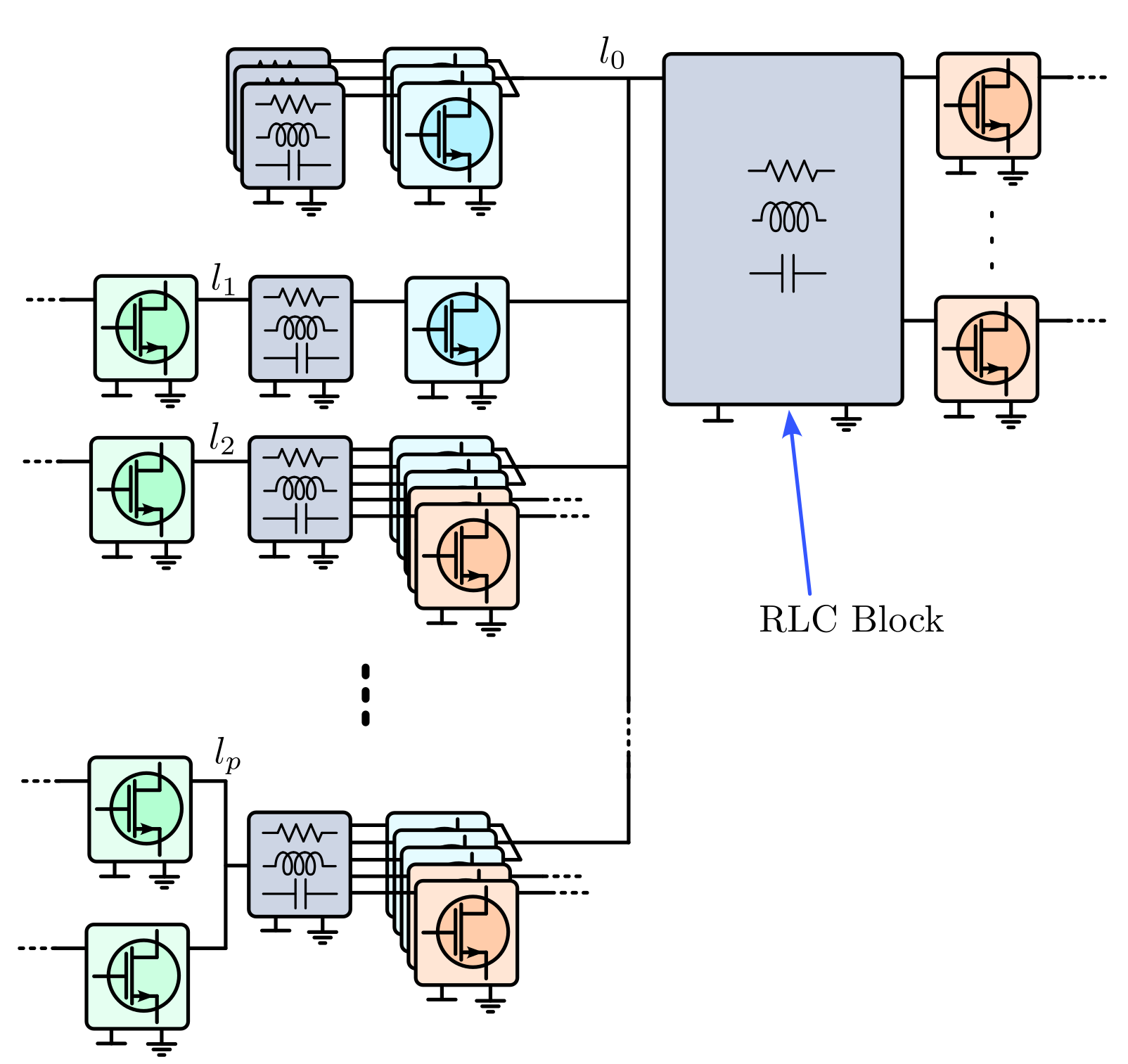}\label{fig: Transistor Network l-th node structure large signal}}
    \hfill
    \subfloat[]{\includegraphics[scale=1]{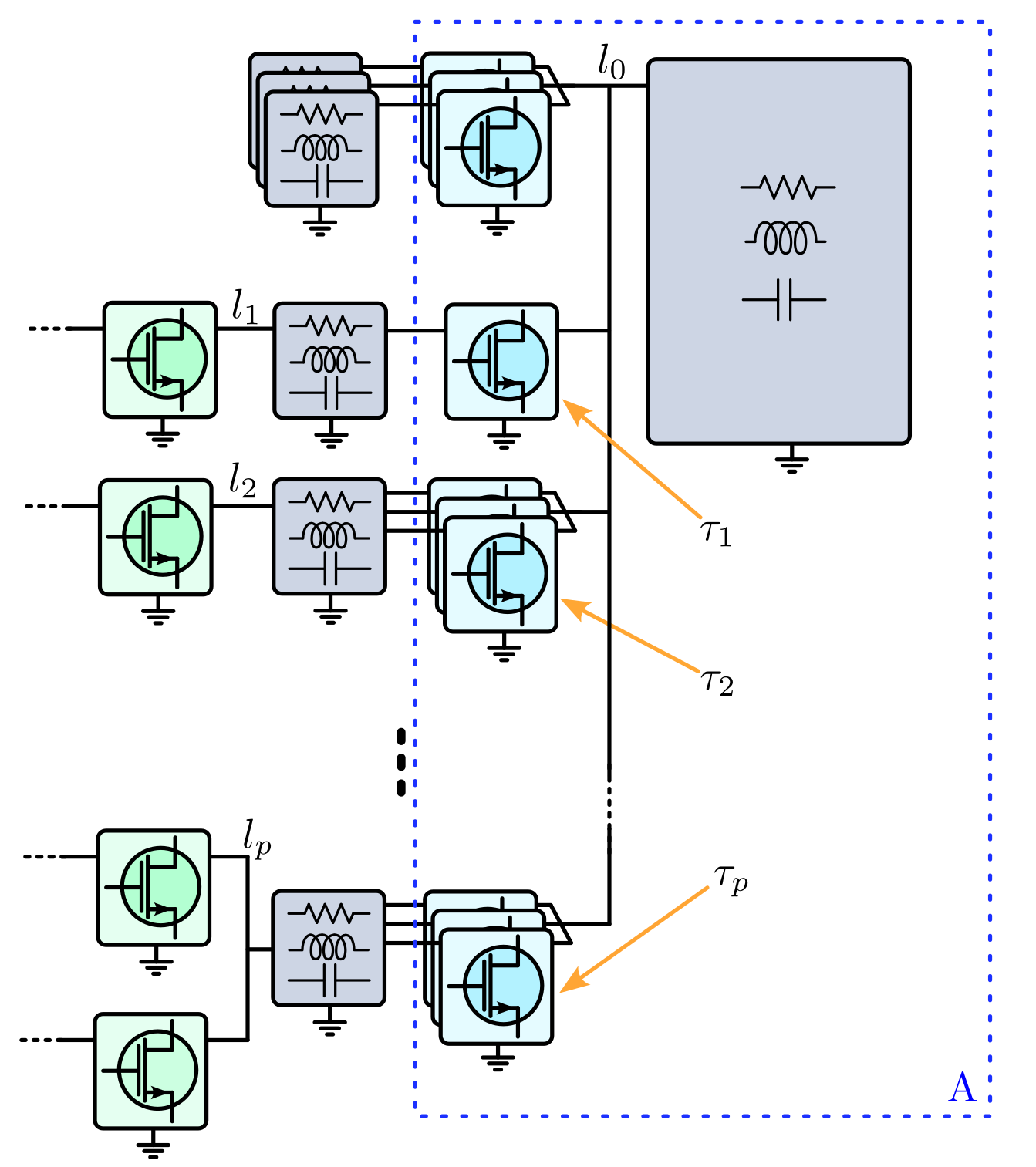}\label{fig: Transistor Network l-th node structure small signal}}
    \hfill
    \caption{Circuit layout around an output node $l_0$; (a) shows detailed connections, (b) shows the transistors and the output nodes that influence the voltage at $l_0$.}
    \label{fig: Transistor Network l-th node structure}
\end{figure}

 Using Assumptions \ref{assump: no rlc between outputs} - \ref{assump: no rlc between same transistor terminals} and properties that follow, it can be inferred that the layout of the considered circuits around an output node $l_0$ can be drawn as shown in Fig. \ref{fig: Transistor Network l-th node structure large signal}. It can also be concluded that the input current of the transistor stages are negligible because of high input impedance. As a result, given the voltages at the nodes $l_1,~l_2,...,l_p$, the amplifier stages that do not have their output ports connected to $l_0$ have no effect in the determination of voltage at $l_0$. Therefore, ignoring the amplifier stages whose output ports are not connected to $l_0,~l_1,...,l_p$, the circuit layout diagram around $l_0$ can be drawn as shown in Fig. \ref{fig: Transistor Network l-th node structure small signal}. Moreover, since we are concerned with only small signal behavior of the circuit we have replaced the power nodes with ground in Fig. \ref{fig: Transistor Network l-th node structure small signal}.
\par With the layout of the circuit described, next we solve Problem \ref{prob: LDIM Model}. We start with the frequency domain description of single amplifier stage and then proceed to analyze larger subcircuits in a hierarchical fashion to show that the voltage at $l_0$ can be determined in a closed form expression that has the form required by Problem \ref{prob: LDIM Model}.
\subsection{MOSFET Amplifier as LDIM}
\begin{figure}[!t]
    \centering
    \subfloat[]{\includegraphics[scale=1]{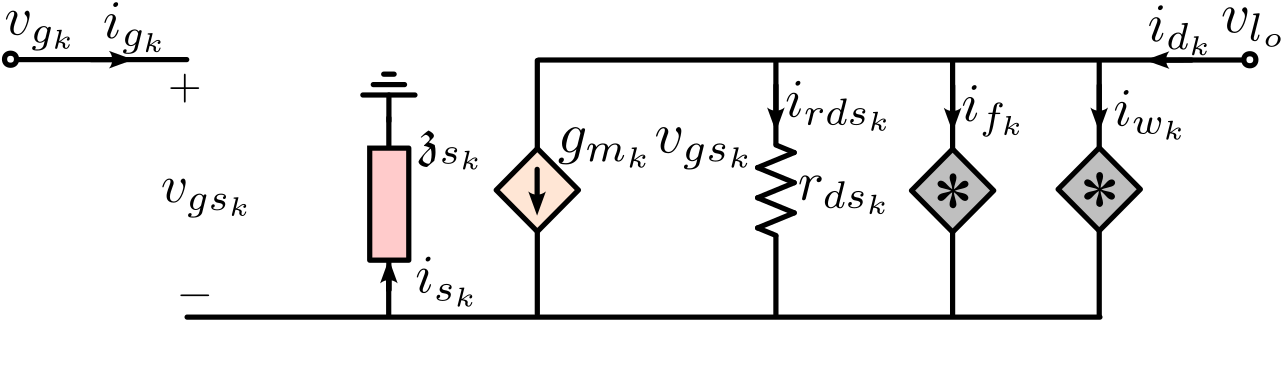}\label{fig: Single MOSFET CS Equivalent}}
    \hfill
    \subfloat[]{\includegraphics[scale=1]{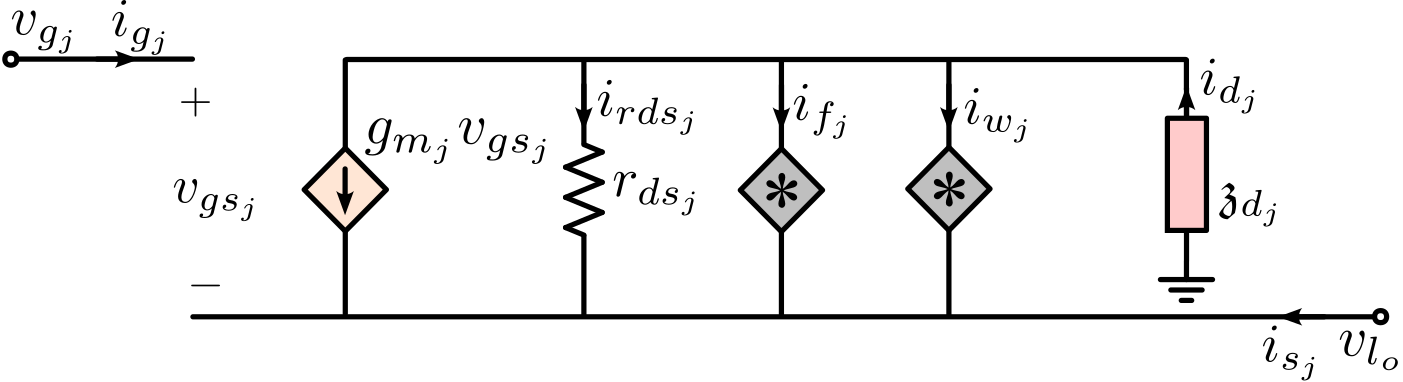}\label{fig: Single MOSFET CD Equivalent}}
    \hfill
    \caption{Small signal noise equivalents circuits of MOSFET: (a) common source mode, (b) common drain mode.}
    \label{fig: Single MOSFET Noise Equivalents}
\end{figure}
\par The first step in determining the voltage at the electrical node $l_0$ is to analyze the subcircuit inside the blue dashed box in Fig. \ref{fig: Transistor Network l-th node structure small signal}. Let this be subcircuit-A as denoted in the diagram of Fig. \ref{fig: Transistor Network l-th node structure small signal}. We start the analysis by discussing the single amplifier stages. As described before, the amplifier stage can be in either in CS or CD configuration. Based on the configuration the small signal model of the amplifier stage can be drawn as shown in Fig. \ref{fig: Single MOSFET Noise Equivalents}. Let the $k^{th}$ transistor in subcircuit-A be in CS configuration with the small signal equivalent model as shown in Fig. \ref{fig: Single MOSFET CS Equivalent}. s-domain (Laplace domain) analysis of the circuit results in the following expression for the drain current of the amplifier stage: 
\begin{align}
    I_{d_k}(s)=&\frac{g_{m_k}}{M_{\alpha_k}(s)} V_{g_k}(s) + \frac{1}{r_{ds_k}M_{\alpha_k}(s)} V_{l_o}(s) \nonumber\\
    &- \frac{1}{M_{\alpha_k}(s)}(I_{f_k}(s)+I_{w_k}(s)).\label{eq: I_o for CS Stage}\\
    \mbox{Here,}&\nonumber\\
    M_{\alpha_k}(s) &:= 1+g_{m_k}\mathfrak{Z}_{s_k}(s)+\frac{\mathfrak{Z}_{s_k}(s)}{r_{ds_k}},
\end{align}
where $\mathfrak{Z}_{s_k}(s)$ is the $s$-domain representation of the impedance $\mathfrak{z}_{s_k}$ in Fig. \ref{fig: Single MOSFET CS Equivalent}.
\par Similarly, if the $j^{th}$ transistor in subcircuit-A is in CD configuration, then the small signal equivalent model is as shown in Fig. \ref{fig: Single MOSFET CD Equivalent}. Further, the source current can be expressed as
\begin{align}
    I_{s_j}(s)=& \frac{g_{m_j}}{M_{\beta_j}(s)}V_{g_j}(s) + \frac{g_{m_j}+\frac{1}{r_{ds_j}}}{M_{\beta_j}(s)} V_{l_o}(s)\nonumber\\
    &- \frac{1}{M_{\beta_j}(s)}(I_{f_j}(s)+I_{w_j}(s)).\label{eq: I_o for CD Stage}\\
    \mbox{Here, }M_{\beta_j}(s) &:= 1+\frac{\mathfrak{Z}_{s_j}(s)}{r_{ds_j}}.
\end{align}
\begin{figure}[!t]
    \centering
    \includegraphics[scale=1]{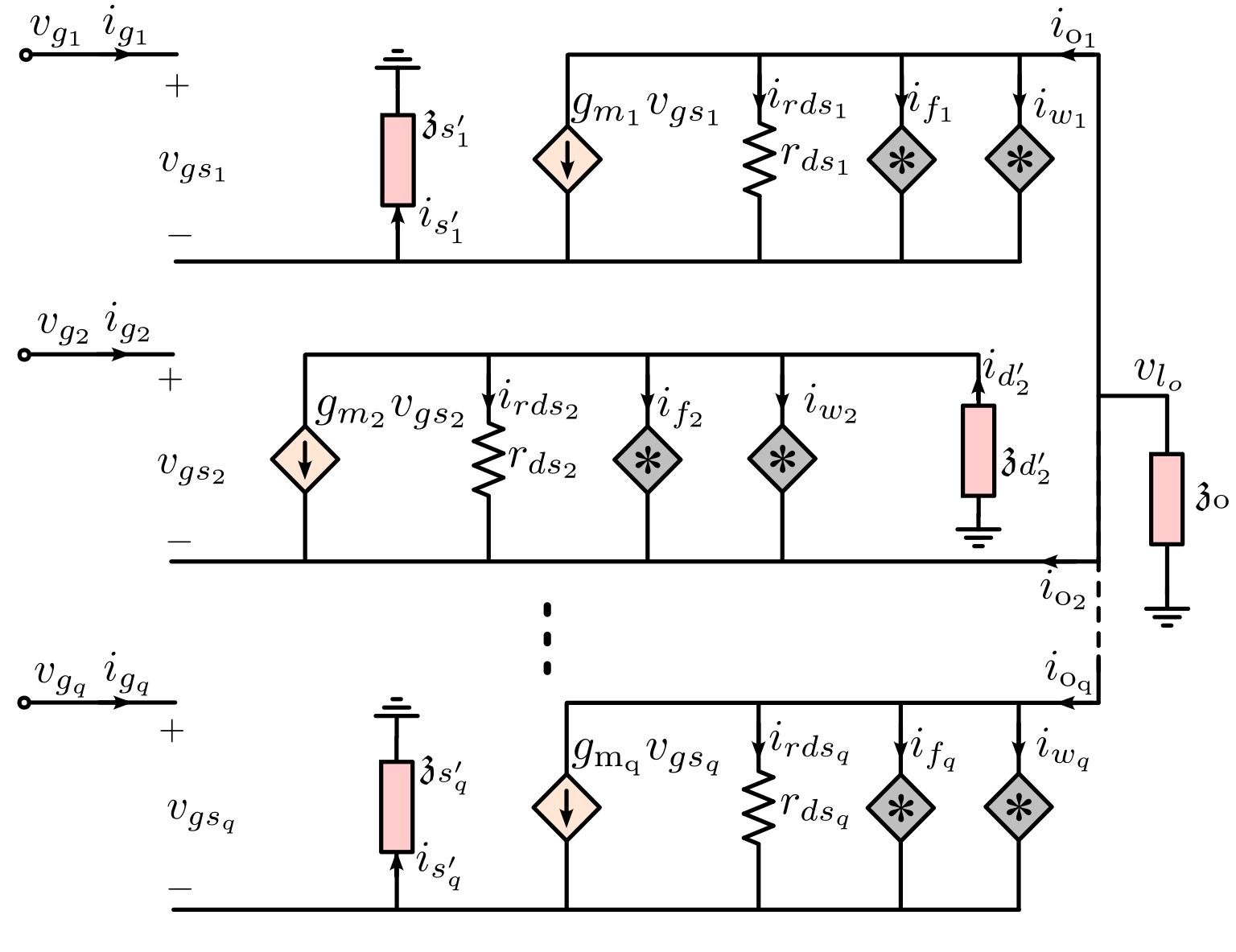}
    \caption{Small signal noise equivalent of multiple FETs in a common source and common drain mode with output connected together.}
    \label{fig: n FET small signal equivalent}
\end{figure}
\par With single amplifier stage described above, we next consider the entire subcircuit-A of Fig. \ref{fig: Transistor Network l-th node structure small signal}. Observe that by replacing the RLC block with its equivalent impedance $(\mathfrak{z}_o)$, the small signal noise equivalent model of subcircuit-A can be drawn as shown in Fig. \ref{fig: n FET small signal equivalent}. As described before, the transistor stages in the subcircuit can be either in CS or CD configuration. We partition the transistor stages into two categories, based on which configuration they are in, by using the following definitions:
\begin{align}
    Q_d^{(l_0)}&:=\{k|\mbox{transistor } k \mbox{ in } A \mbox{ is in CS configuration}\},\nonumber\\
    Q_s^{(l_0)}&:=\{k|\mbox{transistor } k \mbox{ in } A \mbox{ is in CD configuration}\}\mbox{.}\nonumber
\end{align}
\par Observe that by s-domain analysis of the small signal equivalent circuit in Fig. \ref{fig: n FET small signal equivalent}, the voltage at the output node $l_0$ can be written as:
\begin{align}
    V_{l_0}(s)=-\mathfrak{Z}_o(s)\sum_{k\in Q_d^{(l_0)}}I_{d_k}(s)-\mathfrak{Z}_o(s)\sum_{k\in Q_s^{(l_0)}}I_{s_k}(s).\label{eq: V_l0 in terms of I_o}
\end{align}
\par By substituting the expressions for $I_{d_k}(s)$ and $I_{s_k}(s)$ from \eqref{eq: I_o for CS Stage} and \eqref{eq: I_o for CD Stage} into \eqref{eq: V_l0 in terms of I_o} and doing algebraic manipulations, we get
\begin{align}
    V_{l_0}(s)=\sum_{k\in Q_d^{(l_0)}\cup Q_s^{(l_0)}}H_k(s)V_{g_k}(s)+\varepsilon_{l_0}(s),\label{eq: V_l0 in terms of V_gk}
\end{align}
\par where $V_{g_k}(s)$ is the gate voltage of transistor $k$, and $H_k(s)$ is a transfer function with expression 
\begin{align}
    H_k(s):=&\frac{C_k(s)}{1+\sum_{j\in Q_d^{(l_0)}}T_j(s)+\sum_{j\in Q_d^{(l_0)}}S_j(s)},\label{eq: Expression of H_k}\\
    \mbox{with }C_k(s)&:=\begin{cases}
        -\frac{\mathfrak{Z}_o(s)g_{m_k}}{M_{\alpha_k}(s)} & \mbox{if } k\in Q_d^{(l_0)}\\
        -\frac{\mathfrak{Z}_o(s)g_{m_k}}{M_{\beta_k}(s)} & \mbox{if } k\in Q_s^{(l_0)},
    \end{cases}\\
    T_j(s)&:=\frac{\mathfrak{Z}_o(s)}{r_{ds_j}M_{\alpha_j}(s)},\\
    S_j(s)&:=\mathfrak{Z}_o(s)\frac{g_{m_j}+\frac{1}{r_{ds_j}}}{M_{\beta_j}(s)}.
\end{align}
\par $\varepsilon_{l_0}(s)$ in \eqref{eq: V_l0 in terms of V_gk} is a combination of the noise sources of all the transistors in subcircuit-A as shown below:
\begin{align}
    \varepsilon_{l_0}(s)&:=\sum_{k\in Q_d^{(l_0)}\cup Q_s^{(l_0)}}P_k(s)\left(I_{f_k}(s)+I_{w_k}(s)\right),\label{eq: Expression of epsilon_k}\\
    \mbox{where }P_k&:=\frac{D_k(s)}{1+\sum_{j\in Q_d^{(l_0)}}T_j(s)+\sum_{j\in Q_d^{(l_0)}}S_j(s)},\\
    D_k(s)&:=\begin{cases}
        -\frac{\mathfrak{Z}_o(s)}{M_{\alpha_k}(s)} & \mbox{if } k\in Q_d^{(l_0)}\\
        -\frac{\mathfrak{Z}_o(s)}{M_{\beta_k}(s)} & \mbox{if } k\in Q_s^{(l_0)}.
    \end{cases}
\end{align}

\begin{figure}[!t]
    \centering
    \includegraphics[scale=1]{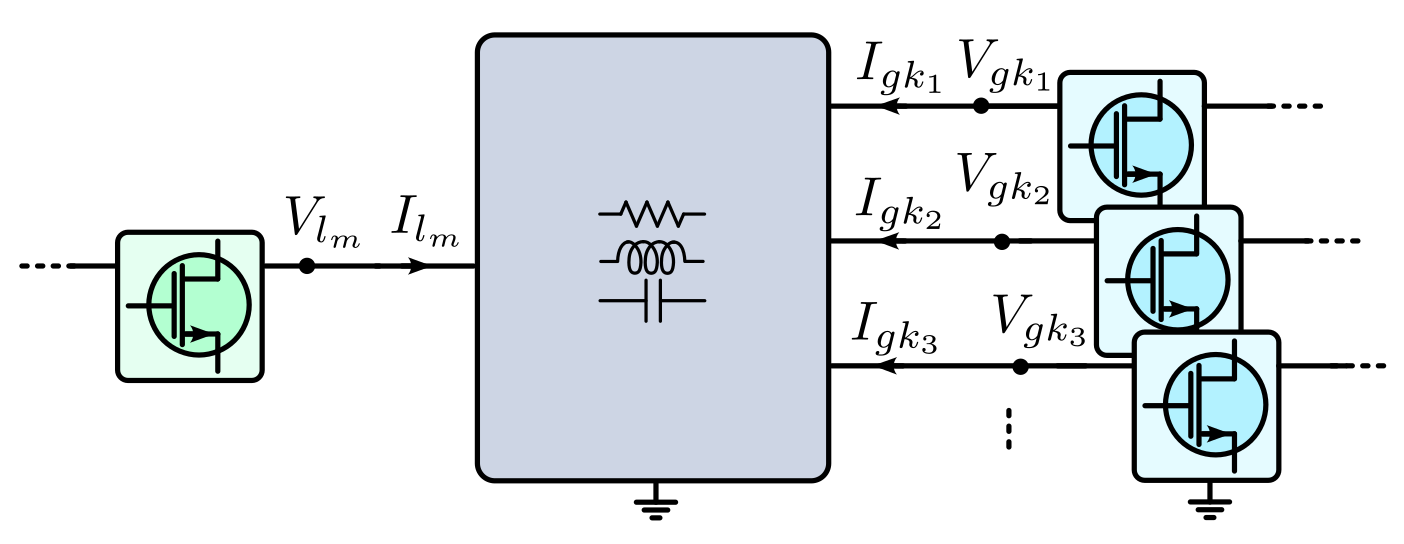}
    \caption{An RLC block with input and output ports shown.}
    \label{fig: RLC Block}
\end{figure}
\par As a logical next step in solving Problem \ref{prob: LDIM Model}, we need to find the gate voltages of the transistors in suncircuit-A for Fig. \ref{fig: Transistor Network l-th node structure small signal} in terms of the output node voltage of the other transistors. To this end, we need to examine the RLC blocks in the circuit. Observe that an RLC block can be connected to at most one output node as described earlier. Thus an RLC block of the circuit can be drawn as in Fig. \ref{fig: RLC Block}. Moreover, since the amplifier stages have high input impedance, all the ports on the right hand side of the RLC block can be approximated to be open. Such RLC networks are well studied in literature (see also \cite{fialkow_RLC_analysis}). Next we categorize the transistor stages in subcircuit-A based on where there inputs are connected by using the definition: $\tau_j:=\{$set of transistors in subcircuit-A that are connected to the RLC block at output node $l_j\}$. Observe that, these are disjoint sets because the input of a transistor stage can not be connected to RLC blocks of two separate output nodes, as that would violate Assumption \ref{assump: no rlc between outputs}. According to results presented in \cite{fialkow_RLC_analysis}, the gate voltages can be derived in terms of the voltage of the output nodes as 
\begin{align}
    V_{g_k}(s)=\begin{cases}
        Z_k(s)V_{l_1}(s) & \forall k\in\tau_1,\\
        Z_k(s)V_{l_2}(s) & \forall k\in\tau_2,\\
        \vdots & \vdots\\
        Z_k(s)V_{l_p}(s) & \forall k\in\tau_p,\\
        0 & \mbox{Otherwise.}
    \end{cases}\label{eq: expression of V_gk}
\end{align}
\par Here, $Z_k(s)$ are the transfer functions of the RLC blocks. Substituting $V_{g_k}(s)$ from \eqref{eq: expression of V_gk} into \eqref{eq: V_l0 in terms of V_gk}, we obtain
\begin{align}
    V_{l_0}(s)=&\left(\sum_{k\in \tau_1}H_k(s)Z_k(s)\right)V_{l_1}(s)+...\nonumber\\
    +&\left(\sum_{k\in \tau_p}H_k(s)Z_k(s)\right)V_{l_p}+\varepsilon_{l_0}(s)\\
    \implies V_{l_0}(s)&=\mathcal{H}_{l_ol_1}(s)V_{l_1}(s)+...\nonumber\\
    &+\mathcal{H}_{l_ol_p}(s)V_{l_p}(s)+\varepsilon_{l_0}(s),\label{eq: V_l in tems of output voltages}\\
    \mbox{where } &\mathcal{H}_{l_ol_m}(s):=\sum_{k\in \tau_m}H_k(s)Z_k(s).
\end{align}
\par Thus we have expressed the voltage of output node $l_0$ in terms of the voltages of other output nodes and a noise term. Similar analysis can be performed at all the $n$ output nodes to obtain $n$ such linear equations. Then the $n$ linear equations can be written in matrix form as 
\begin{align}
    V(s)=\mathcal{H}(s)V(s)+\varepsilon(s),
\end{align}
where ${V}(s)$ is a $n\times1$ vector of voltages of the output nodes, $\mathcal{H}(s)$ is a $n\times n$ transfer matrix, and $\varepsilon(s)$ is a $n\times1$ vector of noise processes of the transistors.
\par Using s-domain to z-domain transforms such as Bilinear transform, we can obtain 
\begin{align}
    \hat{V}(z)=\hat{\mathcal{H}}(z)\hat{V}(z)+\hat{\varepsilon}(z).
\end{align}
\par On the unit circle, $|z|=1$, this model can be written as
\begin{align}
    \hat{V}(\omega)=\hat{\mathcal{H}}(\omega)\hat{V}(\omega)+\hat{\varepsilon}(\omega),
\end{align}
\par which is precisely the expression in \eqref{eq: voltage LDIM}.

\par Next, we examine the cross-power spectral density of $\varepsilon_{l}(\cdot)$ and $\varepsilon_{m}(\cdot)$ where, $l\neq m$. Using \eqref{eq: Expression of epsilon_k} we can write
\begin{align}
    \hat{\varepsilon}_{l}(\omega)&=\sum_{k\in Q_d^{(l)}\cup Q_s^{(l)}}\hat{P}_k(\omega)\left(\hat{I}_{f_k}(\omega)+\hat{I}_{w_k}(\omega)\right),\label{eq: Expression of epsilon_k in z-domain}
\end{align}
which implies
\begin{align}
    \hat{\varepsilon}_l(\omega)&=\mathcal{P}_l(\omega)\mathfrak{I}_l(\omega),
\end{align}
\par where
\begin{align}
    \mathcal{P}_l(\omega)&:=\left[\begin{array}{cccc}
        \hat{P}_{k_1}(\omega) & \hat{P}_{k_1}(\omega) & ... & \hat{P}_{k_t}(\omega)
    \end{array}\right],\\
    \mathfrak{I}_l(\omega)& :=\left[\begin{array}{ccccc}
        \hat{I}_{f_{k_1}}(\omega) & \hat{I}_{w_{k_1}}(\omega) & ... & \hat{I}_{w_{k_t}}(\omega)
    \end{array}\right]^T,\\
    \mbox{and }&\{k_1,~k_2,~...,~k_t\}=Q_d^{(l)}\cup Q_s^{(l)}. 
\end{align}

\par Using the assumption of independence of transistor noise processes (Assumption \ref{assump: Independent noise process}) we conclude that $\Phi_{\mathfrak{I}_l\mathfrak{I}_m}(\omega)=0$. Therefore, $\Phi_{{\varepsilon}_l{\varepsilon}_m}(\omega)=\mathcal{P}_l(\omega)\Phi_{\mathfrak{I}_l\mathfrak{I}_m}(\omega)\mathcal{P}_m^*(\omega)=0$. Thus, we have shown that the PSD matrix is diagonal. This completes the solution to Problem \ref{prob: LDIM Model}.

\begin{remark}
    The reader might get the impression that to be able to express the circuit as an LDIM one needs to measure all the output nodes. Fortunately, one can measure a subset of the output nodes and still be able to express the influence flow as an LDIM. Some instances of such scenarios are shown in later subsections. However, it should be noted that the choice of the measured subset of the output nodes need to be judicious; improper choice of measured subset may lead to inability to obtain an LDIM. The exact conditions for which particular nodes need to be measured to have an LDIM representation is a subject of future research.
\end{remark}


\subsection{Influence Flow Model of Cascode Amplifier Network}

\begin{figure}[!t]
    \centering
    \includegraphics[scale=1]{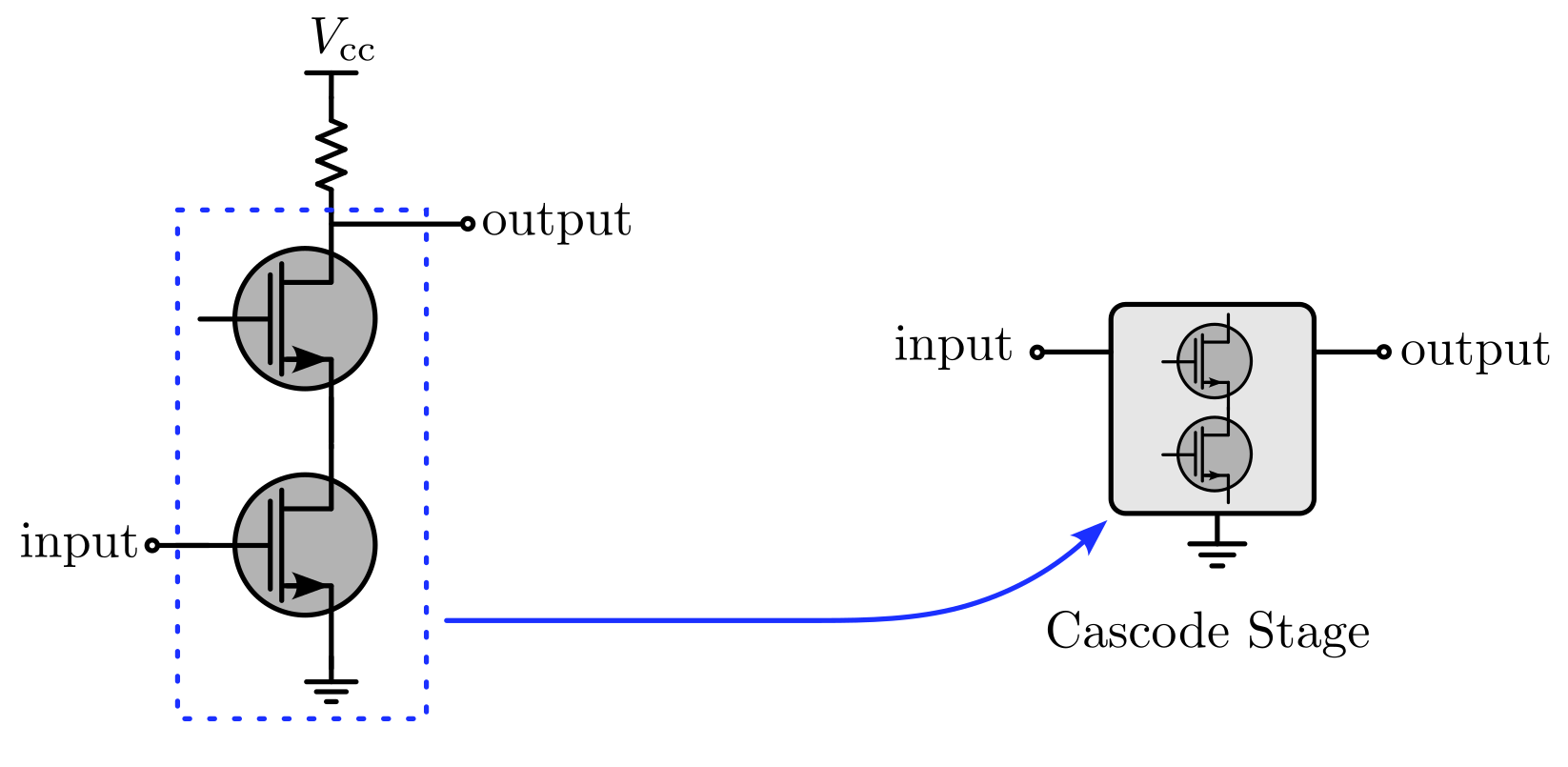}
    \caption{A cascode amplifier with its symbolic representation.}
    \label{fig: Cascode Stage}
\end{figure}

\par Nest we turn our focus to a different type of amplifier, namely, the cascode amplifiers. A \emph{cascode stage} is a combination of a common source transistor with a common gate stage, as shown in Fig. \ref{fig: Cascode Stage} (see also \cite{razaviCMOSIC}). Given Assumptions \ref{assump: no rlc between outputs} and \ref{assump: no rlc between same transistor terminals}, similar analysis as in previous sections can be done to infer that the circuit layout around an output node will have a structure as in Fig. \ref{fig: Transistor Network l-th node structure large signal} with the \emph{transistor stages} replaced by the \emph{cascode stages}. Therefore, a similar analysis can be performed to derive the underlying influence model that has the form shown in \eqref{eq: voltage LDIM}.


\subsection{Example}

\begin{figure}[!t]
    \centering
    \subfloat[]{\includegraphics[scale=1]{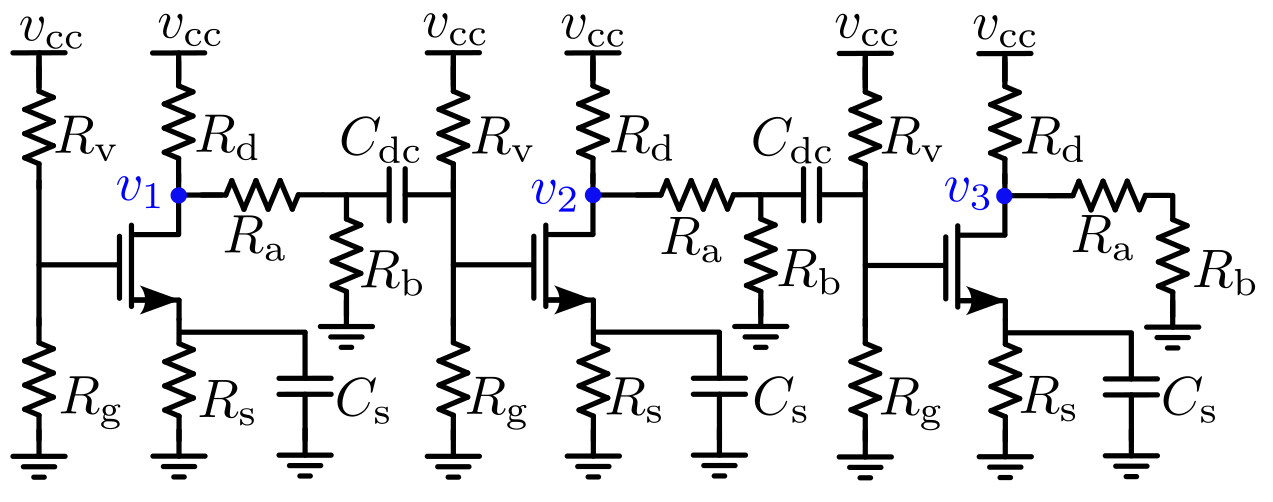}\label{fig: MOSFET LDIM Example Circuit}}
    \hfill
    \subfloat[]{\includegraphics[scale=1]{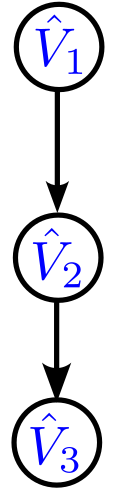}\label{fig: Graphical represasentation of example circuit}}
    \caption{(a) An example circuit to illustrate modeling approach (b) graphical representation of LDIM of the circuit.}
    \label{fig: MOSFET LDIM Example Circuit and its graph}
\end{figure}

\par Next we present an example to illustrate the modeling approach. Consider the circuit in Fig. \ref{fig: MOSFET LDIM Example Circuit}. The LDIM of the circuit can be derived as:
\begin{align}\label{eq: example circuit LDIM}
    \left[\begin{array}{c}
         \hat{V}_1(\omega)\\
         \hat{V}_2(\omega)\\
         \hat{V}_3(\omega)
    \end{array}\right]=\left[\begin{array}{ccc}
        0 & 0 & 0 \\
        \hat{\mathcal{H}}_{21}(\omega) & 0 & 0 \\
        0 & \hat{\mathcal{H}}_{32}(\omega) & 0
    \end{array}\right]&\left[\begin{array}{c}
         \hat{V}_1(\omega)\\
         \hat{V}_2(\omega)\\
         \hat{V}_3(\omega)
    \end{array}\right]\nonumber\\+&\left[\begin{array}{c}
         \hat{\varepsilon}_1(\omega)\\
         \hat{\varepsilon}_2(\omega)\\
         \hat{\varepsilon}_3(\omega)
    \end{array}\right].
\end{align}

\par A graphical representation of the LDIM is given in Fig. \ref{fig: Graphical represasentation of example circuit}. Observe that only two of the entries of the transfer function matrix are non-zero in \eqref{eq: example circuit LDIM}, namely $\hat{\mathcal{H}}_{21}(\omega)$ and $\hat{\mathcal{H}}_{32}(\omega)$. It captures the fact that $v_3(\cdot)$ is influenced only by $v_2(\cdot)$, and $v_2(\cdot)$ is influenced only by $v_1(\cdot)$ in the circuit as shown in Fig. \ref{fig: Graphical represasentation of example circuit}, which is a direct consequence of the connectivity structure of the circuit. Another interpretation of the generative structure could be in terms of the signal flow structure of the circuit. Observe that the generative graph accurately captures the signal flow paths in the considered circuit.

\begin{remark}
    Observe that in the above example, one does not need to measure all the nodes for the underlying model to be an LDIM. For example, one could exclude the measurement of any one of the nodes and the resulting model would still be an LDIM.
\end{remark}


\section{Simulation Results}\label{sec: Simulation Results}

\begin{table}
    \caption{MOSFET Circuit Simulation Parameters}
    \vspace{-0.35cm}
    \label{Table: Simulation Circuit Paramater}
    \begin{center}
    \begin{tabular}{c|c||c|c}
        \hline
        \textbf{Parameters} & \textbf{Value} & \textbf{Parameters} & \textbf{Value} \\
        \hline
        $R_{v}$ & 40 K$\Omega$ & $R_{g}$ & 8 K$\Omega$\\
        \hline
        $R_{d}$ & 0.6 K$\Omega$ & $R_{s}$ & 0.1 K$\Omega$\\
        \hline
        $R_{a}$ & 42 K$\Omega$ & $R_{b}$ & 1 K$\Omega$\\
        \hline
        $C_{s}$ & 1 $\mu$F & $C_{dc}$ & 1 $\mu$F\\
        \hline
        $v_{cc}$ & 12 V & $R_{o}$ & 43 K$\Omega$\\
        \hline
        $R_{c}$ & 0.3 K$\Omega$ & $R_{m}$ & 200 K$\Omega$\\
        \hline
        $R_{e}$ & 100 K$\Omega$ & $R_{f}$ & 75 K$\Omega$\\
        \hline
        $R_{h}$ & 5 K$\Omega$ & $R_{p}$ & 4 K$\Omega$\\
        \hline
        $R_{t}$ & 0.5 K$\Omega$ & $C_{o}$ & 10 $\mu$F\\
        \hline
        $C_{t}$ & 10 $\mu$F & $C_{u}$ & 10 $\mu$F\\
        \hline
        $v_{ss}$ & 15 V & $R_{r}$ & 42 K$\Omega$\\
        \hline
    \end{tabular}
    \end{center}
    \vspace{-0.7cm}
\end{table}
\par In this section we present simulation results. Here we use the reconstruction algorithm described in Section \ref{sec: preliminaries} on time-series voltage measurement data collected from transient noise simulations performed in Cadence Spectre circuit simulator. 
\subsection{Simulation Result-1}
\subsubsection{Complete Measurement}
\begin{figure}[!t]
    \centering
    \subfloat[]{\includegraphics[scale=1]{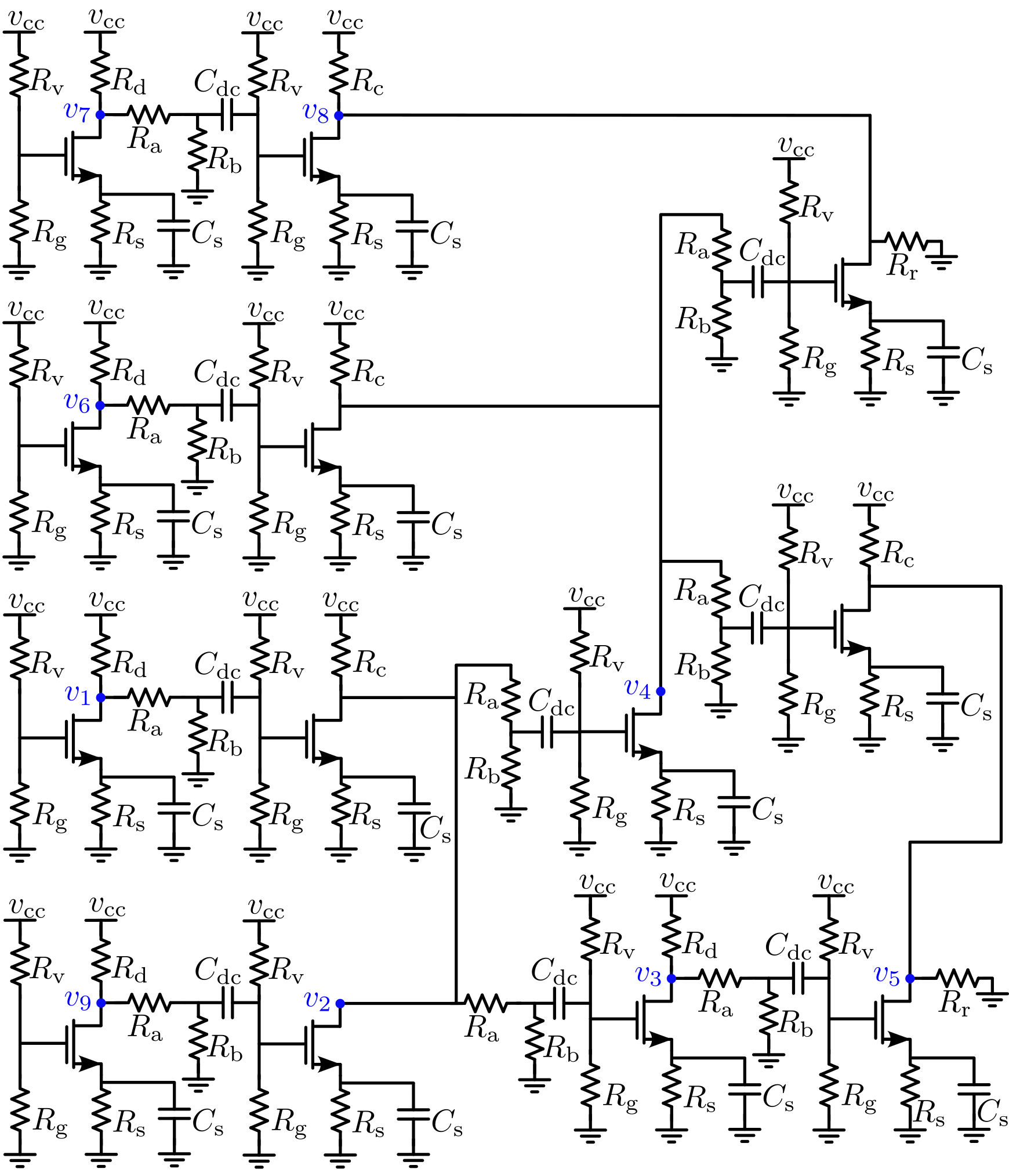}\label{fig: 9 nodes Simulated network}}
    \hfill
    \subfloat[]{\includegraphics[scale=1]{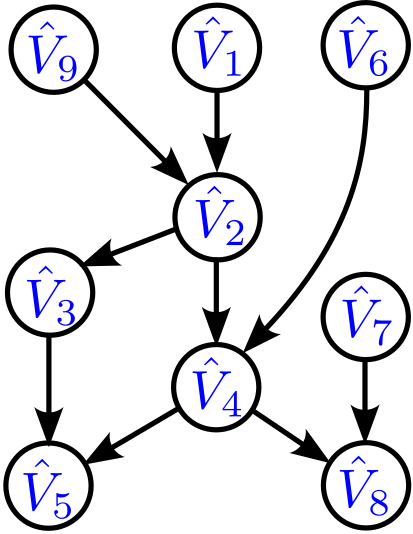}\label{fig: 9 nodes generative graph}}
    \hfill
    \subfloat[]{\includegraphics[scale=1]{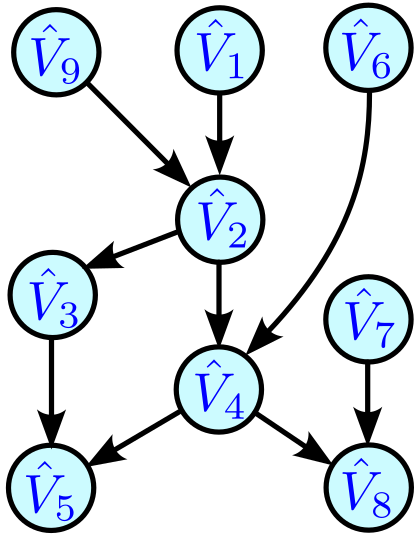}\label{fig: 9 nodes reconstructed graph}}
    \hfill
    \subfloat[]{\includegraphics[scale=1]{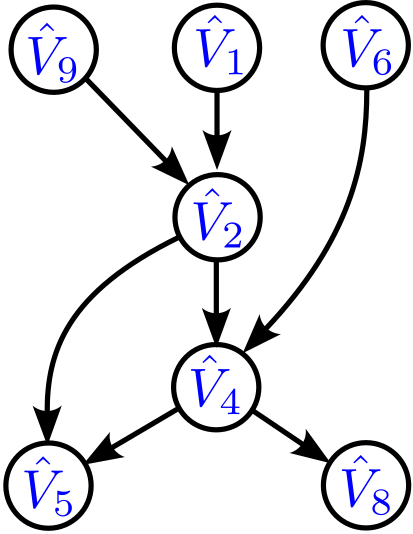}\label{fig: 7 nodes generative graph}}
    \hfill
    \subfloat[]{\includegraphics[scale=1]{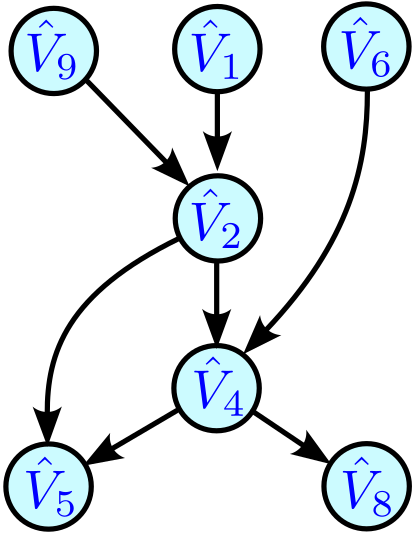}\label{fig: 7 nodes reconstructed graph}}
    \hfill
    \caption{(a) An amplifier circuit with $9$ output nodes connected in a mesh like network, (b) generative structure for the circuit, (c) reconstructed graph, (d) generative structure for the circuit with partial measurements, (e) reconstructed graph with partial measurements.}
    \label{fig: 9 nodes meshed MOSFET network diagrams}
\end{figure}
\par The first set of results that we present are for a nine stage amplifier where the amplifier stages are connected in a meshed fashion as shown in Fig. \ref{fig: 9 nodes Simulated network}. It can be inferred using modeling techniques described in Section-\ref{sec: LDIM of Transistor Amplifiers} that the generative structure is of the form shown in Fig. \ref{fig: 9 nodes generative graph}. A transient noise simulation of the circuit in Fig. \ref{fig: 9 nodes Simulated network} was performed in Cadence Spectre. The maximum and minimum noise frequency chosen for the simulation were $1$ MHz and $1$ kHz respectively. A level-1 spice model of an n-channel MOSFET with $V_{to}=1.21411V,~\lambda=0.000503783V^{-1},~k_p=0.800568A/V^{-1},~C{gso}=8.25824\times10^{-8}F/m,~C_{gdo}=1.56921\times10^{-8}F/m$ along with the circuit parameters given in Table \ref{Table: Simulation Circuit Paramater} were used in the simulation. It is to be noted that even though we assumed that passive elements are noiseless during the modeling step, in the simulations we consider all the resistive elements to be noisy to capture a more realistic scenario. $850000$ samples of time-series measurements for each of the voltages $v_1(\cdot),...,v_9(\cdot)$ were recorded. The data was then used in the reconstruction algorithm described in Section-\ref{sec: preliminaries}. The Wiener filters were estimated using the method described in \cite{veedu_wiener_pc}. The estimated Wiener filters were used to verify the Wiener separation condition with a threshold of $\rho=0.05$. The result obtained from the reconstruction algorithm is shown in Fig. \ref{fig: 9 nodes reconstructed graph}. Observe that in this case the PC algorithm was able to reconstruct the complete generative structure. Observe that the reconstructed graph accurately captures the signal flow paths in the circuit of Fig. \ref{fig: 9 nodes Simulated network}.

\subsubsection{Partial Measurement}

\par Next we showcase that not all the output nodes have to be measured for the presented modeling and reconstruction approaches to be feasible. To that end we exclude the measurements of $v_3(\cdot)$ and $v_7(\cdot)$ of the circuit in Fig. \ref{fig: 9 nodes Simulated network}. In that case the generative structure takes the form shown in Fig. \ref{fig: 7 nodes generative graph}. The measured partial data was used in the reconstruction algorithm with similar parameters as in case of the complete measurement experiment to arrive at the result shown in Fig. \ref{fig: 7 nodes reconstructed graph}. It can be observed that the reconstructed graphs is accurate. This substantiates our claim that not all nodes have to be measured; however, one has to judiciously choose which nodes to exclude.  


\subsection{Simulation Results-2}

\begin{figure}[!t]
    \centering
    \subfloat[]{\includegraphics[scale=1]{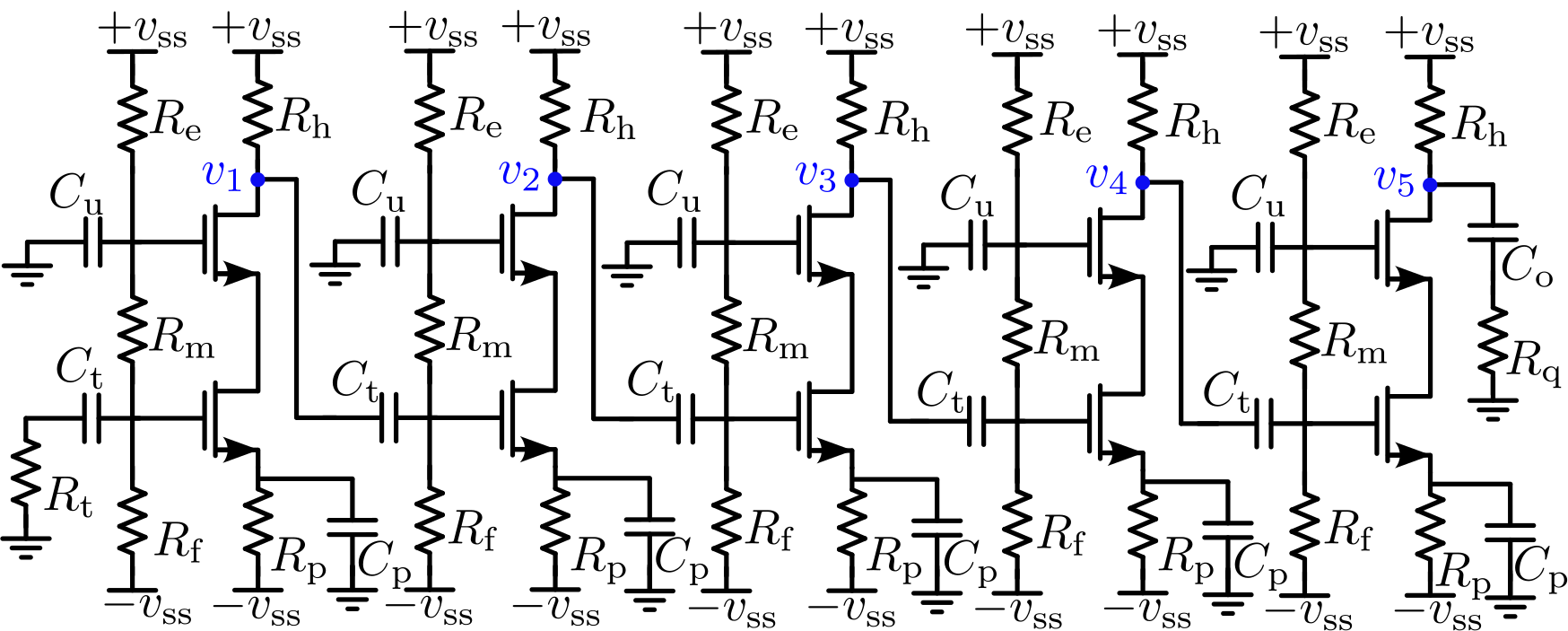}\label{fig: 5 Stage Simulated Cascode Chain}}
    \hfill
    \subfloat[]{\includegraphics[scale=1]{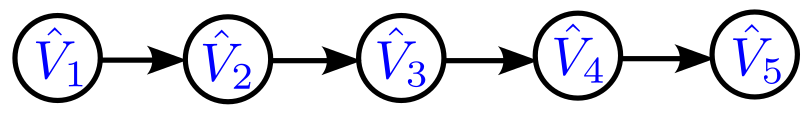}\label{fig: 5 Stage Cascode Chain generative Graph}}
    \hfill
    \subfloat[]{\includegraphics[scale=1]{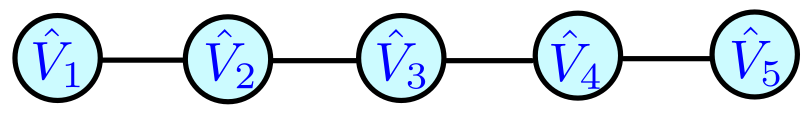}\label{fig: 5 Stage Cascode Chain Reconstructed Graph}}
    \hfill
    \caption{(a) A network of $5$ cascode stages in a chain like layout, (b) generative structure for the circuit, (c) reconstructed graph.}
    \label{fig: 5 Stage Cascode Chain Diagrams}
\end{figure}

\par Next we present results for a $5$ stage cascode amplifier as shown in Fig. \ref{fig: 5 Stage Simulated Cascode Chain}. The generative structure of the circuit is shown in Fig. \ref{fig: 5 Stage Cascode Chain generative Graph}. A transient noise simulation of the circuit in Fig. \ref{fig: 5 Stage Simulated Cascode Chain} was performed in Cadence Spectre with maximum and minimum noise frequency of $500$ kHz and $1$ kHz respectively. A level-1 spice model of an n-channel MOSFET with $V_{to}=0.7V,~\lambda=0.04V^{-1},~k_p=330\times10^{-6}A/V^{-1}$ along with the circuit parameters given in Table \ref{Table: Simulation Circuit Paramater} were used in the simulation. $480000$ samples of measurements of $v_1(\cdot),...,v_5(\cdot)$ were recorded and then used in the reconstruction algorithm. The Wiener filters were estimated using the measured data samples that were further used to verify the Wiener separation condition with a threshold of $\rho=0.064$. The result obtained from the reconstruction algorithm is shown in Fig. \ref{fig: 5 Stage Cascode Chain Reconstructed Graph}. It can be observed that the skeleton of the generative structure is reconstructed correctly, which captures the interconnection structure of the amplifier stages.

\subsection{Simulation Results-3}
\par The third set of results presented are for a BJT amplifier network. BJT amplifiers share structural and operational similarity with MOSFET amplifiers. Therefore, the modeling and reconstruction principles developed in this article can be applied to such BJT amplifier circuits.
\subsubsection{Complete Measurement}

\begin{figure}[!t]
    \centering
    \subfloat[]{\includegraphics[scale=1]{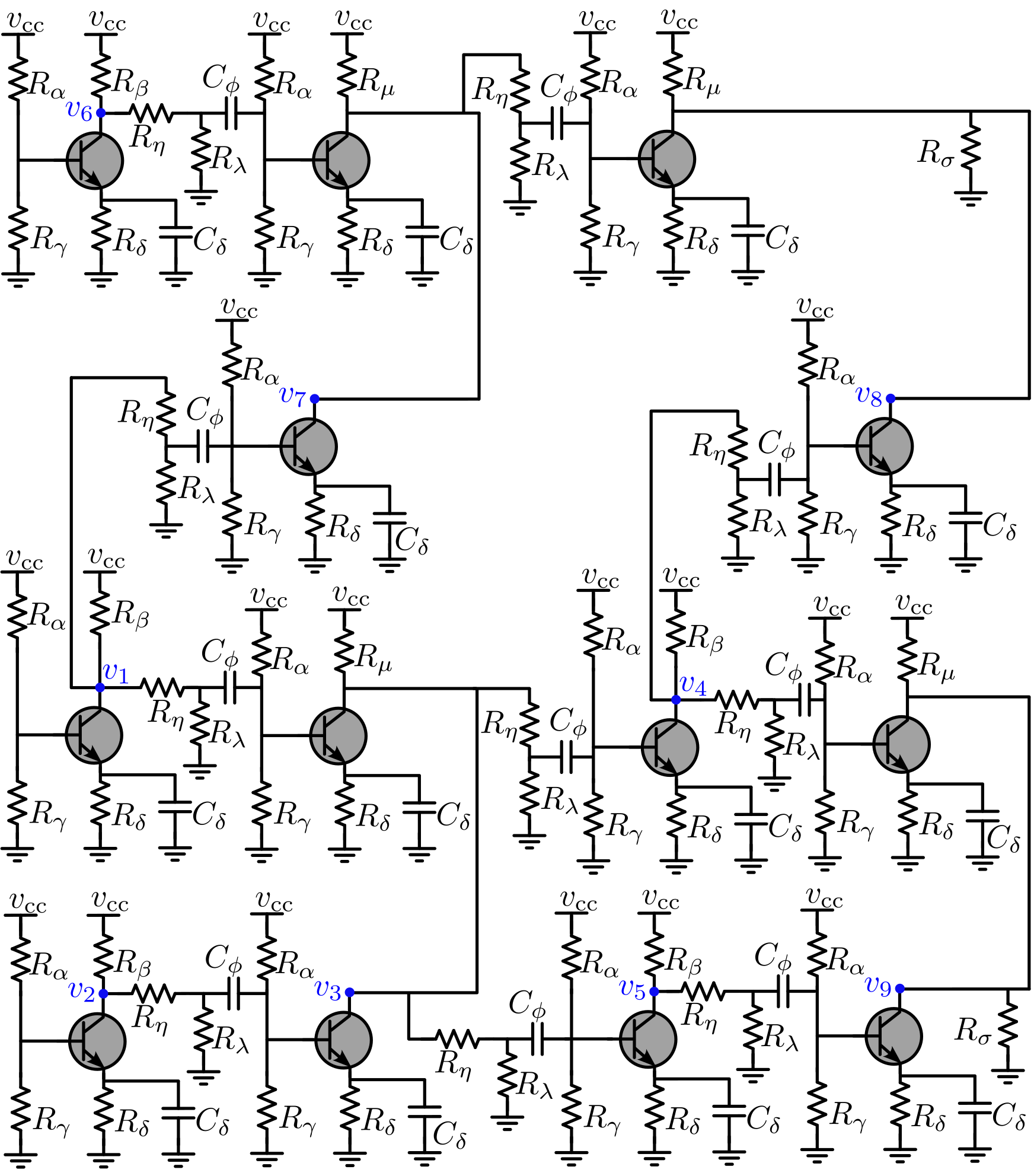}\label{fig: 9 Nodes Simulated BJT Circuit}}
    \hfill
    \subfloat[]{\includegraphics[scale=1]{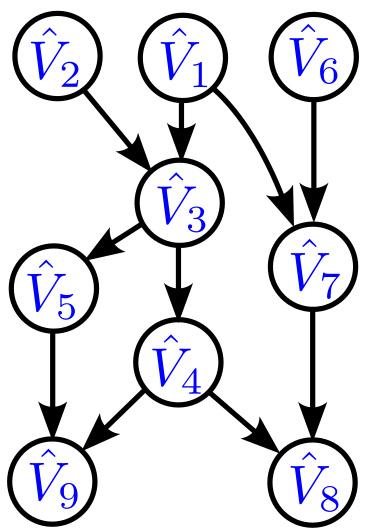}\label{fig: 9 Nodes BJT Circuit generative Graph}}
    \hfill
    \subfloat[]{\includegraphics[scale=1]{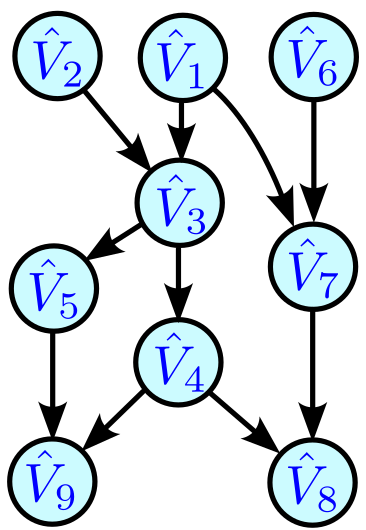}\label{fig: 9 Nodes BJT Circuit reconstructed Graph}}
    \hfill
    \subfloat[]{\includegraphics[scale=1]{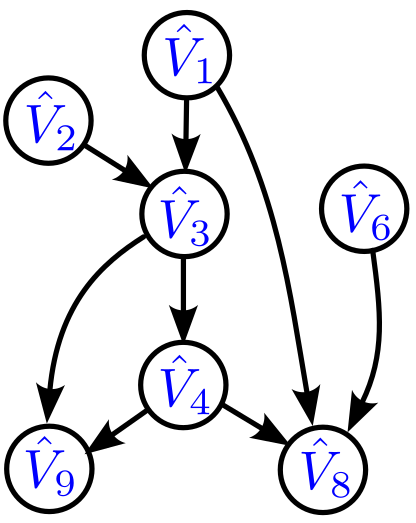}\label{fig: 7 Nodes BJT Circuit generative Graph}}
    \hfill
    \subfloat[]{\includegraphics[scale=1]{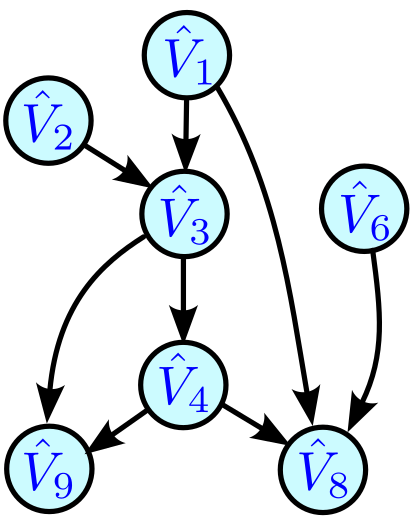}\label{fig: 7 Nodes BJT Circuit reconstructed Graph}}
    \hfill
    \caption{(a) A network of BJT amplifier stages, (b) generative structure for the circuit, (c) reconstructed graph, (d) generative structure for the circuit with partial measurements, (e) reconstructed graph with partial measurements.}
    \label{fig: 9 Nodes BJT Circuit Diagrams}
\end{figure}

\begin{table}
    \caption{BJT Circuit Simulation Parameters}
    \vspace{-0.35cm}
    \label{Table: BJT Simulation Circuit Paramater}
    \begin{center}
    \begin{tabular}{c|c||c|c}
        \hline
        \textbf{Parameters} & \textbf{Value} & \textbf{Parameters} & \textbf{Value} \\
        \hline
        $R_\mathrm{\alpha}$ & 100 K$\Omega$ & $R_\mathrm{\beta}$ & 22 K$\Omega$\\
        \hline
        $R_\mathrm{\gamma}$ & 22 K$\Omega$ & $R_\mathrm{\delta}$ & 4.7 K$\Omega$\\
        \hline
        $R_\mathrm{\mu}$ & 11 K$\Omega$ & $R_\mathrm{\eta}$ & 22 K$\Omega$\\
        \hline
        $R_\mathrm{\lambda}$ & 22 K$\Omega$ & $C_\mathrm{\phi}$ & 1 $\mu$F\\
        \hline
        $C_\mathrm{\delta}$ & 1 pF & $v_\mathrm{cc}$ & 12 V\\
        \hline
        $R_\mathrm{\sigma}$ & 1 G$\Omega$ &  &\\
        \hline
    \end{tabular}
    \end{center}
    \vspace{-0.7cm}
\end{table}

\par We present results for a BJT amplifier circuit as shown in Fig. \ref{fig: 9 Nodes Simulated BJT Circuit}. The generative structure of the circuit is shown in Fig. \ref{fig: 9 Nodes BJT Circuit generative Graph}. A transient noise simulation of the circuit in Fig. \ref{fig: 9 Nodes Simulated BJT Circuit} was performed in Cadence Spectre with maximum and minimum noise frequency of $1$ MHz and $1$ kHz respectively. A spice model of a commercial BJT (MMBT2222ATT1G) along with the circuit parameters given in Table \ref{Table: BJT Simulation Circuit Paramater} were used in the simulation. In this case also all the resistive elements were considered to be noisy. $799999$ samples of measurements of $v_1(\cdot),...,v_9(\cdot)$ were recorded and then used in the reconstruction algorithm. The Wiener filters were estimated using measured data samples that were further used to verify the Wiener separation condition with a threshold of $\rho=0.033$. The result obtained from the reconstruction algorithm is shown in Fig. \ref{fig: 9 Nodes BJT Circuit reconstructed Graph}. It can be seen that the generative structure is correctly reconstructed, which captures the interconnection and signal flow structure.

\subsubsection{Partial Measurement}

\par Next we exclude $v_5(\cdot)$ and $v_7(\cdot)$ from our measurements. The collected data was used in the reconstruction algorithm with same parameter settings as in the case of the complete measurement of the BJT circuit. The generative graph and the reconstructed graph for the partial measurement case are shown in Fig. \ref{fig: 7 Nodes BJT Circuit generative Graph} and Fig. \ref{fig: 7 Nodes BJT Circuit reconstructed Graph}. It can be observed that the reconstruction was accurate.

\section{Experimental Results}\label{sec: Experimental Results}

\begin{figure}[!t]
    \centering
    \includegraphics[scale=0.9]{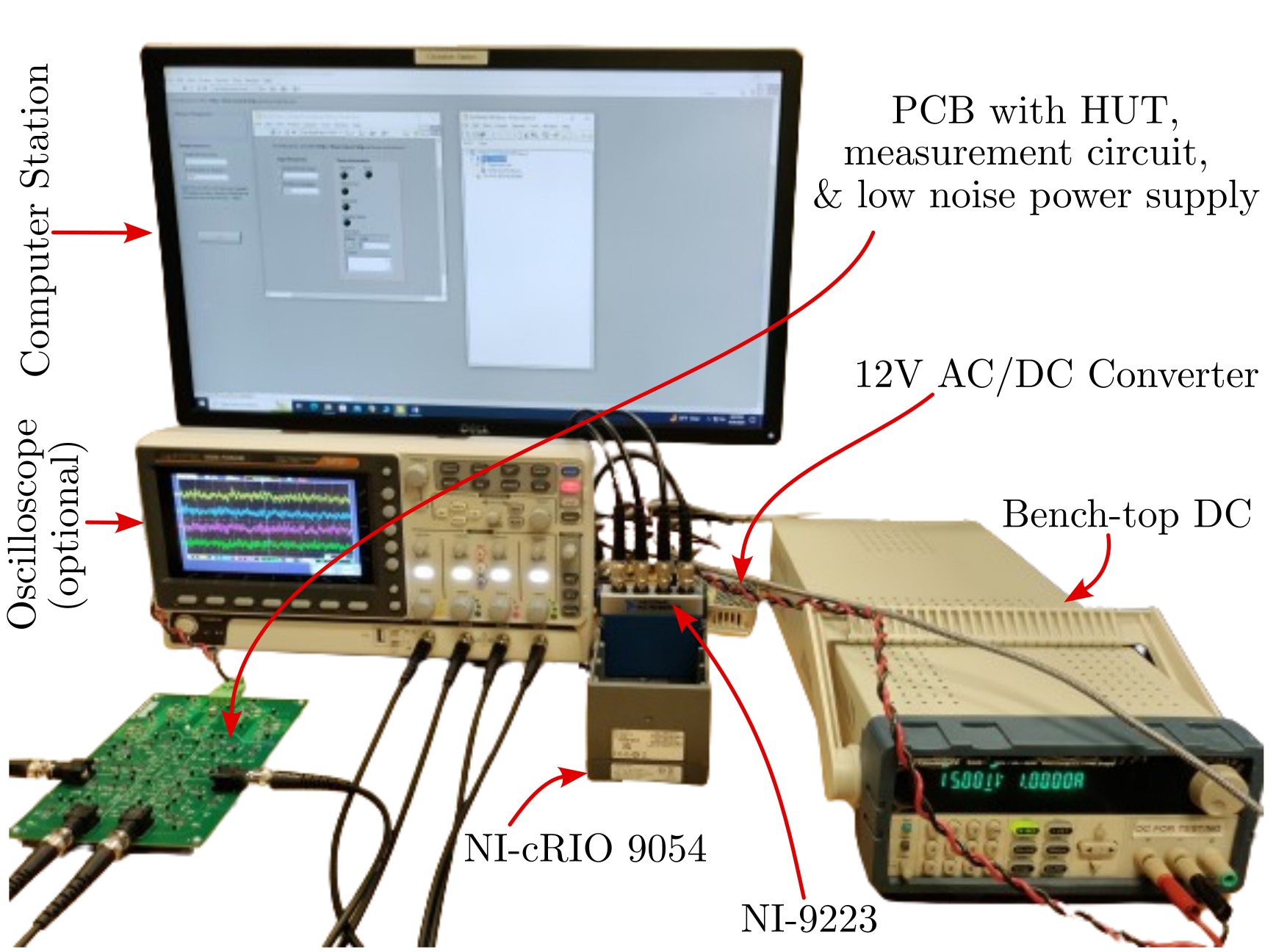}
    \caption{Experimental setup for data acquisition and network reconstruction}
    \label{fig: Experimental Setup}
\end{figure}

\par To further demonstrate the efficacy of the methods presented in this article, we present results obtained for physical hardware prototypes developed in laboratory. The experimental setup is shown in Fig. \ref{fig: Experimental Setup}. The setup comprise of the following subsystems: 

\begin{itemize}
    \item a computer station with LabVIEW 2021 and Python,
    
    \item an NI-cRIO 9054 equipped with dual Intel Atom E3805 1.33 GHz cores, and Xilinx Artix-7 A100T FPGA,
    
    \item an NI-9223 module with following specifications: 4 $\pm 10$V isolated analog input channels, $16$-bit simultaneous sampling of all channels at maximum $1$MS/s rate, 

    \item a 600W Multi-Range 150V/10A bench-top DC Power Supply (Model 9206) from BK Precision,

    \item a 12V 50W AC/DC Converter (LRS-50-12) from Mean Well USA Inc,

    \item low noise instrumentation amplifier (AD$8421$) based custom made measurement circuits,

    \item hardware under test (HUT) which was either a MOSFET amplifier circuit or a BJT amplifier circuit as discussed in subsequent sections.
\end{itemize} 

\begin{figure}[!t]
    \centering
    \includegraphics[scale=0.95]{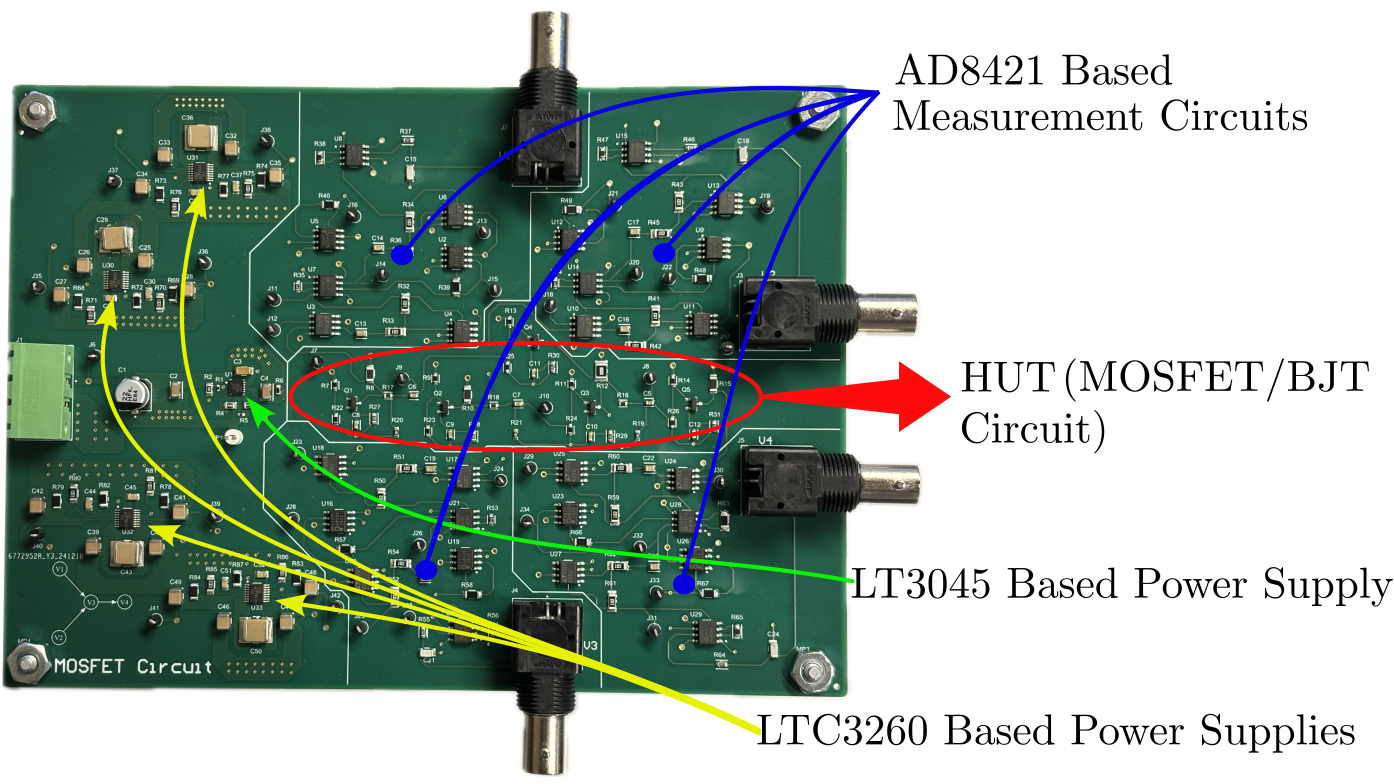}
    \caption{PCB with HUT, low noise power supply, and measurement circuit}
    \label{fig: PCB Experimental Setup}
\end{figure}

\par The output node voltages of the HUT was measured using the AD$8421$ based measurement circuit. The measured signals were acquired in the form of time-series measurements at $1$MS/s using the NI-9223 module, NI-cRIO, and LabVIEW. The LRS-50-12 power supply was used to provide power to the NI-cRIO. The power for the AD$8421$ base measurement circuit was provided using a LTC$3260$ based $12$V low noise charge pump circuit. To avoid complications due to power supply noise, a LT$3045$ based $12$V low noise power supply was designed for the HUT. The LT$3045$ and LTC$3260$ based power supplies were energized using the DC power supply from BK Precision. A picture of the PCB implementing the HUT, low noise power supply, and the measurement circuits is shown in Fig. \ref{fig: PCB Experimental Setup}. Two such PCB boards were developed; one for a MOSFET based HUT and another for a BJT based HUT.

\subsection{MOSFET Amplifier}

\begin{figure}[!t]
    \centering
    \subfloat[]{\includegraphics[scale=1]{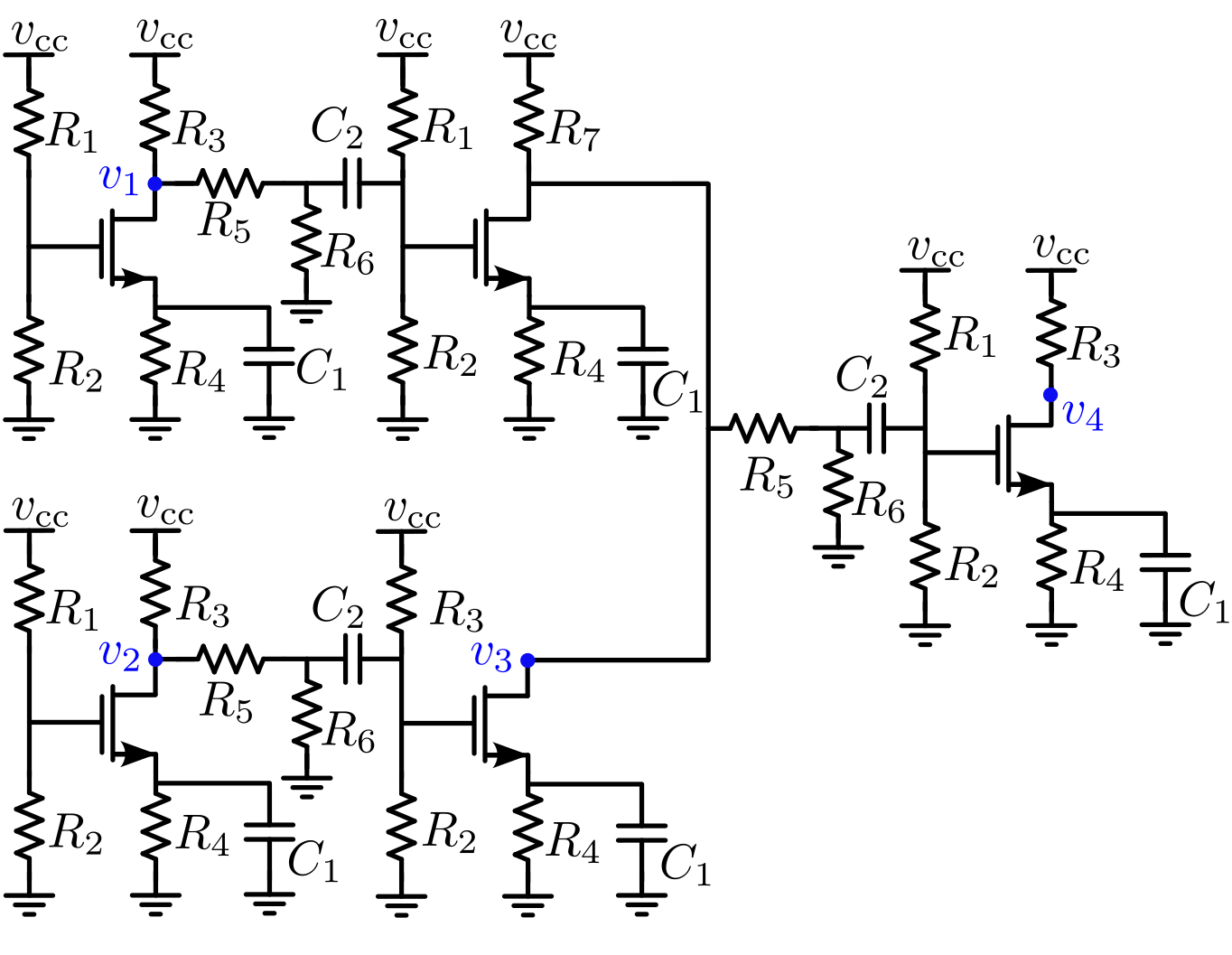}\label{fig: MOSFET Hardware Circuit}}
    \hfill
    \subfloat[]{\includegraphics[scale=1]{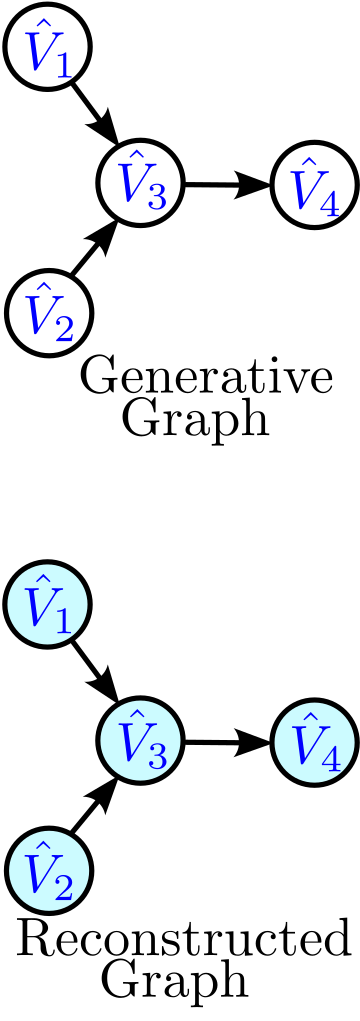}\label{fig: MOSFET Hardware reconstructed graph}}
    \hfill
    \caption{(a) MOSFET Amplifier circuit under test, (b) its generative graph, and reconstructed graph.}
    \label{fig: MOSFET Hardware Diagrams}
\end{figure}

\begin{table}
    \caption{MOSFET HUT Parameters}
    \vspace{-0.35cm}
    \label{Table: MOSFET HUT Paramater}
    \begin{center}
    \begin{tabular}{c|c||c|c}
        \hline
        \textbf{Parameters} & \textbf{Value} & \textbf{Parameters} & \textbf{Value} \\
        \hline
        $R_\mathrm{1}$ & 32 K$\Omega$ & $R_\mathrm{2}$ & 8 K$\Omega$\\
        \hline
        $R_\mathrm{3}$ & 0.597 K$\Omega$ & $R_\mathrm{4}$ & 0.1 K$\Omega$\\
        \hline
        $R_\mathrm{5}$ & 41.7 K$\Omega$ & $R_\mathrm{6}$ & 3 K$\Omega$\\
        \hline
        $R_\mathrm{7}$ & 0.3 K$\Omega$ & $C_\mathrm{1}$ & 1 $\mu$F\\
        \hline
        $C_\mathrm{2}$ & 1 $\mathrm{\mu}$F & $v_\mathrm{cc}$ & 12 V\\
        \hline
    \end{tabular}
    \end{center}
    \vspace{-0.7cm}
\end{table}

\par A prototype of a multistage MOSFET amplifier as shown in Fig. \ref{fig: MOSFET Hardware Circuit} was developed to verify the ability to reconstruct its topology. The circuit consists of five amplifier stages made with $2$N$7002$E n-channel MOSFETS from Onsemi and circuit components with values as shown in  Table \ref{Table: MOSFET HUT Paramater}. Time-series measurements for $v_1(\cdot),...,v_4(\cdot)$ were recorded using the measurement and data acquisition setup described before. A total of $500000$ data samples at a rate of $1$MS/s were recorded and used in the PC algorithm with Wiener separation test. The Wiener filters were estimated using the measured data samples that were further used to verify the Wiener separation condition with a threshold of $\rho=0.028$. The reconstructed graph from the data is shown in Fig. \ref{fig: MOSFET Hardware reconstructed graph}. It can be observed that the algorithm was able to reconstruct the generative structure correctly.

\subsection{BJT Amplifier}

\begin{figure}[!t]
    \centering
    \subfloat[]{\includegraphics[scale=1]{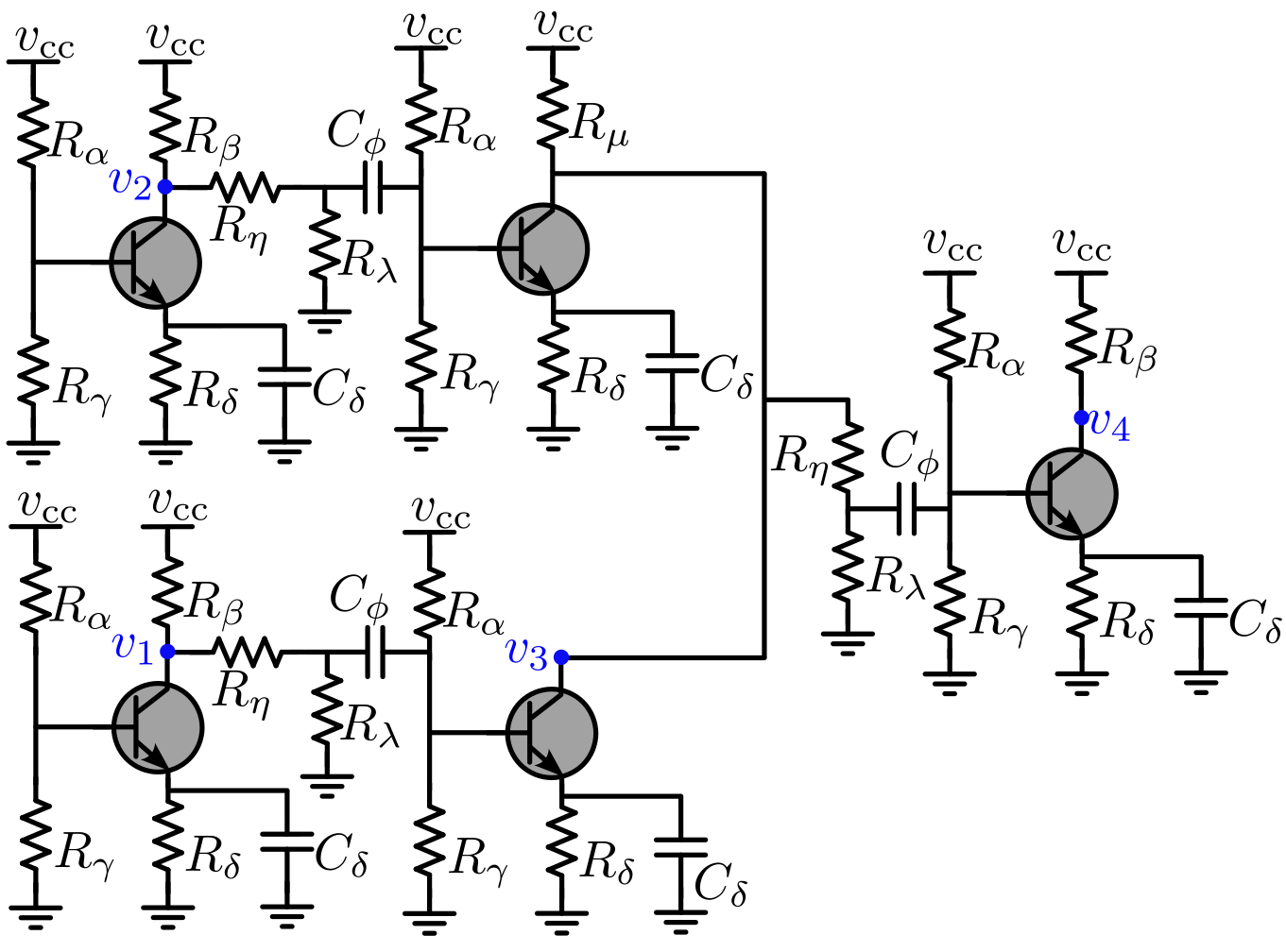}\label{fig: BJT Hardware Circuit}}
    \hfill
    \subfloat[]{\includegraphics[scale=1]{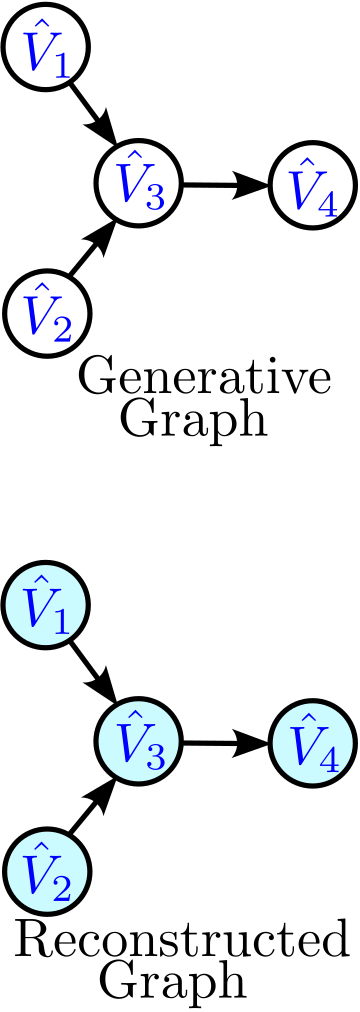}\label{fig: BJT Hardware reconstructed graph}}
    \hfill
    \caption{(a) BJT Amplifier circuit under test, (b) its generative graph, and reconstructed graph.}
    \label{fig: BJT Hardware Diagrams}
\end{figure}

\par We further show the applicability of the reconstruction approach to a real BJT hardware setup. The circuit shown in Fig. \ref{fig: BJT Hardware Circuit} was developed using MMBT$2222$ATT$1$G BJT units from Onsemi and circuit components with values shown in Table \ref{Table: BJT Simulation Circuit Paramater}. Using the hardware setup described before, $500000$ samples of $v_1(\cdot),...,v_4(\cdot)$ were recorded at a rate of $1$MS/s, which was then used in the reconstruction algorithm. The Wiener filters were estimated using the measured data samples that were further used to verify the Wiener separation condition with a threshold of $\rho=0.03$. The generative graph and the reconstructed graph are shown in Fig. \ref{fig: BJT Hardware reconstructed graph}. It can be observed that the algorithm reconstructs the graph correctly.

\section{Application Note}\label{sec: Application Note}

\begin{figure}[!t]
    \centering
    \subfloat[]{\includegraphics[scale=1]{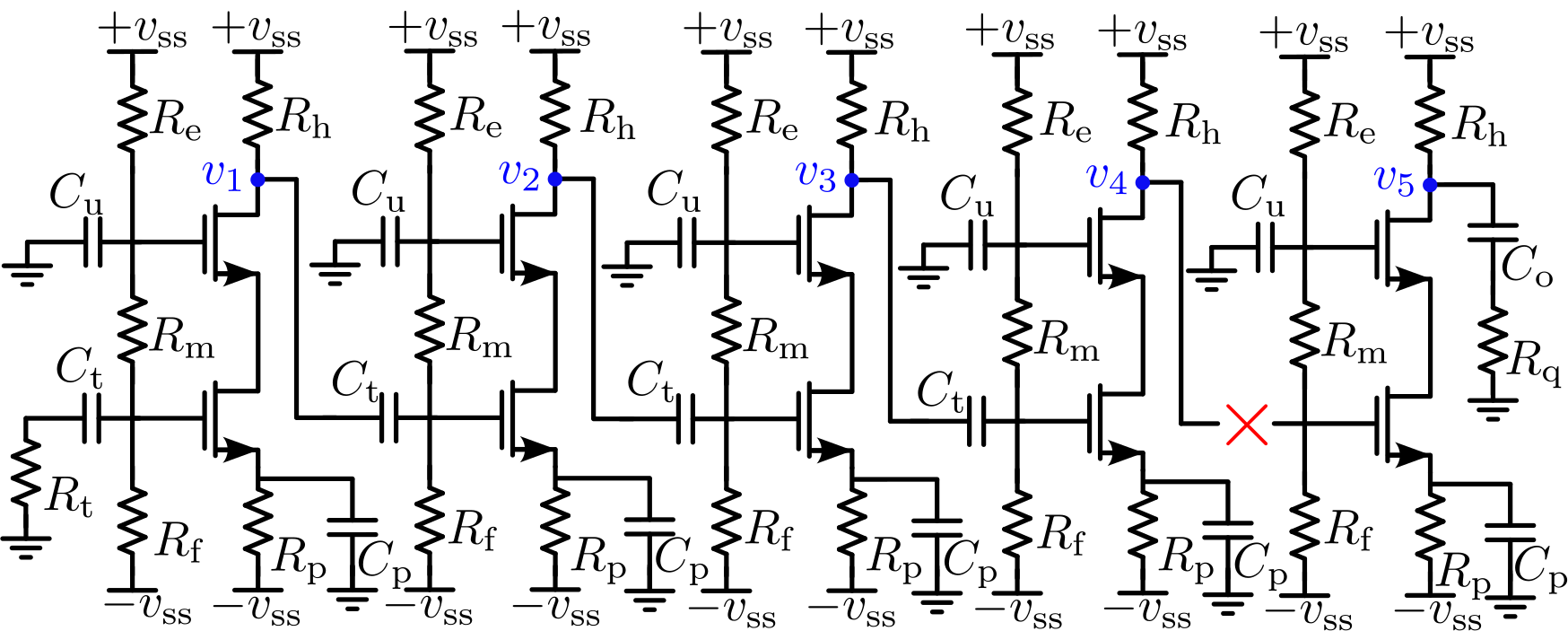}\label{fig: 5 Stage Simulated Cascode Chain with fault}}
    \hfill
    \subfloat[]{\includegraphics[scale=1]{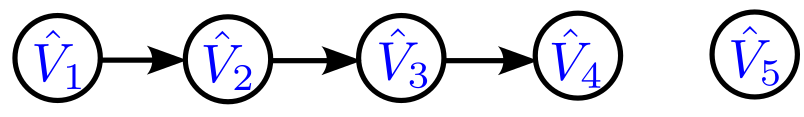}\label{fig: 5 Stage Cascode Chain with fault Generative graph}}
    \hfill
    \subfloat[]{\includegraphics[scale=1]{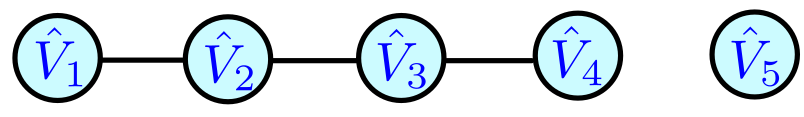}\label{fig: 5 Stage Cascode Chain with fault reconstructed graph}}
    \hfill
    \caption{(a) A network of $5$ cascode stages with fault, (b) generative structure for the circuit, (c) reconstructed graph.}
    \label{fig: 5 Stage Cascode Chain with fault Diagrams}
\end{figure}

\par In this section we present several potential application of the findings in this article. First we show that the reconstruction of network structure of amplifier networks from time-series measurement data can aid in fault diagnosis of the circuits. For example open circuit faults that break some of the signal flow paths in the network can be detected by a simple reconstruction instance using data from the faulty circuit. To further motivate the application an example is shown here. Next it is shown that it can be used for identification of underlying transfer functions of the circuits. Finally, its applicability in evaluation of causal inference algorithms are pointed out.

\par Consider the five-stage cascode amplifier network of Section-\ref{sec: Simulation Results} but this time with an open circuit fault as shown in Fig. \ref{fig: 5 Stage Simulated Cascode Chain with fault}. Intuitively, the fault in the circuit breaks the signal flow path between $\hat{V}_4$ and $\hat{V}_5$ as shown in the generative graph of Fig. \ref{fig: 5 Stage Cascode Chain with fault Generative graph}. Even though a designer may be able to find the fault by doing iterative tests at the different electrical nodes and manually analyzing the measured signals, it might be difficult in a automated scenario to do such iterative tests. Moreover, such manual test might be time consuming and cumbersome. However, reconstruction of the faulty circuit will reveal that the edge between $\hat{V}_4$ and $\hat{V}_5$ is broken, which can help an automated system to readily detect the fault. To verify the hypothesis, a simulation of the faulty circuit of Fig. \ref{fig: 5 Stage Simulated Cascode Chain with fault} was performed in Cadence Spectre with same set up as in case of the five stage cascode circuit in Section \ref{sec: Simulation Results}. The reconstruction result using the data collected from the faulty circuit and algorithm parameters as in case of the five stage cascode network in Section-\ref{sec: Simulation Results} is shown in Fig. \ref{fig: 5 Stage Cascode Chain with fault reconstructed graph}. It can be observed that indeed the edge between $\hat{V}_4$ and $\hat{V}_5$ is not present in the reconstructed graph which is a indication that there is a fault between the two nodes in the circuit. Thus we have shown that network reconstruction can be of great help in fault diagnosis of the circuit.

\par In addition to fault diagnosis, modeling transistor circuits as LDIM also opens up the possibility of efficient data-driven parameter estimation. For example, once the graphical model of the network is identified (or if it is known apriori), it can be used for signal selection to estimate the individual transfer functions of the amplifier stages using the methods described in \cite{materassi_Signal_selection}. The identification of these individual transfer functions can further be used for design verification of such circuits post-production.

\par The ability to model transistor amplifiers as LDIM opens up the possibility for its use in the field of \emph{causal inference} and \emph{network reconstruction} \cite{Peter_clark_book}. The inaccessibility of well-understood real-word datasets makes it challenging to evaluate causal inference algorithms\cite{runge_causeme_1,runge_causeme_2}. However, as shown in this article, the ground truth of influence flow structures of transistor amplifiers can be interpreted correctly in terms of the LDIM and its generative graph. Thus, making it possible to device a platform to effectively evaluate causal inference algorithms on real-world datasets using measurement data from transistor amplifiers.

\section{Conclusion}\label{sec: Conclusion}

\par This article builds bridges between study of transistor amplifiers and theory of networked systems. Firstly, it was established that many transistor amplifiers can be modeled as linear dynamic influence model. Multistage amplifiers with transistors connected in common source, source follower, and cascode configurations were shown to have a underlying signal flow structure that can be captured by an LDIM. As a second contribution, it is shown that network reconstruction tools can be used to reconstruct features of underlying generative structure of transistor amplifiers. Moreover, this article provides directions to use measurement data driven network reconstruction techniques in fault diagnosis applications. Potential application of the findings of this article in another research domain, namely, \emph{causal inference} is noted.

\bibliographystyle{IEEEtran}
\bibliography{bibliography}

\vspace{11pt}

\vspace{-33pt}
\begin{IEEEbiography}[{\includegraphics[width=1.25in,height=1in,clip,keepaspectratio]{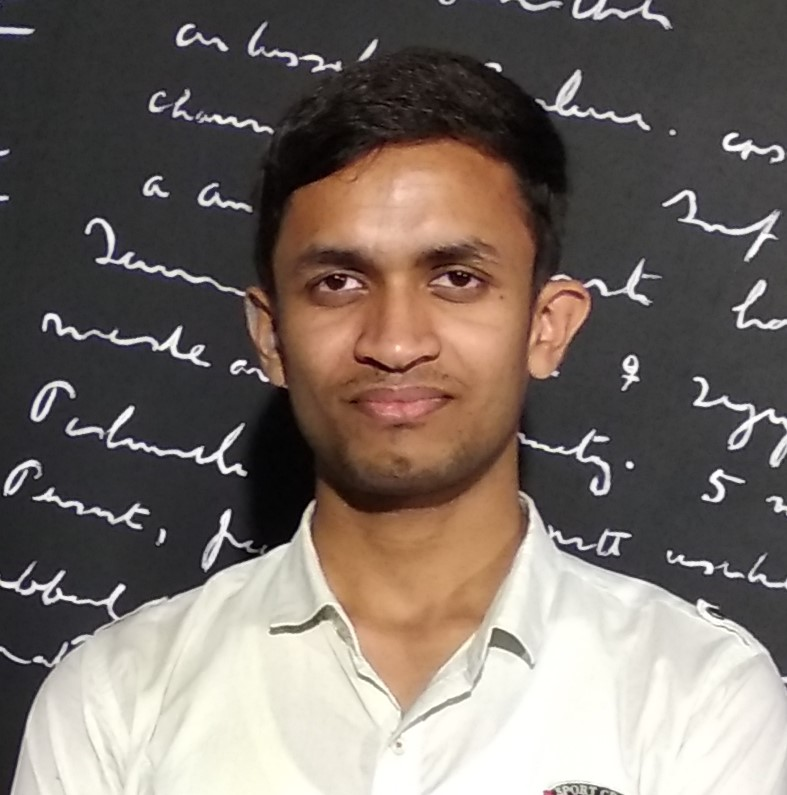}}]{Mohammed Tuhin Rana}
(Graduate Student Member, IEEE) received his B.Tech. degree in electrical engineering from National Institute of Technology Durgapur, West Bengal, India, in 2021, MS degree in electrical and computer engineering from University of Minnesota, MN, USA in 2024, and is currently pursuing a Ph.D. degree in electrical engineering at the University of Minnesota, Twin Cities, USA. His research interests include network reconstruction of dynamic networked systems, control of power electronics, and microgrids.
\end{IEEEbiography}

\begin{IEEEbiography}[{\includegraphics[width=1.25in,height=1in,clip,keepaspectratio]{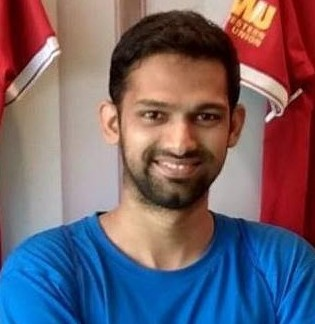}}]{Mishfad Shaikh Veedu} received the M.E. degree in telecommunication engineering from the Indian Institute of Science, Bangalore, India, in 2014 the Ph.D. degree in
electrical engineering with the University of Minnesota, Minneapolis, MN, USA in 2024.
His research interests include optimization,
system identification, estimation theory, information theory, and statistical learning.
\end{IEEEbiography}

\begin{IEEEbiography}[{\includegraphics[width=1.25in,height=1in,clip,keepaspectratio]{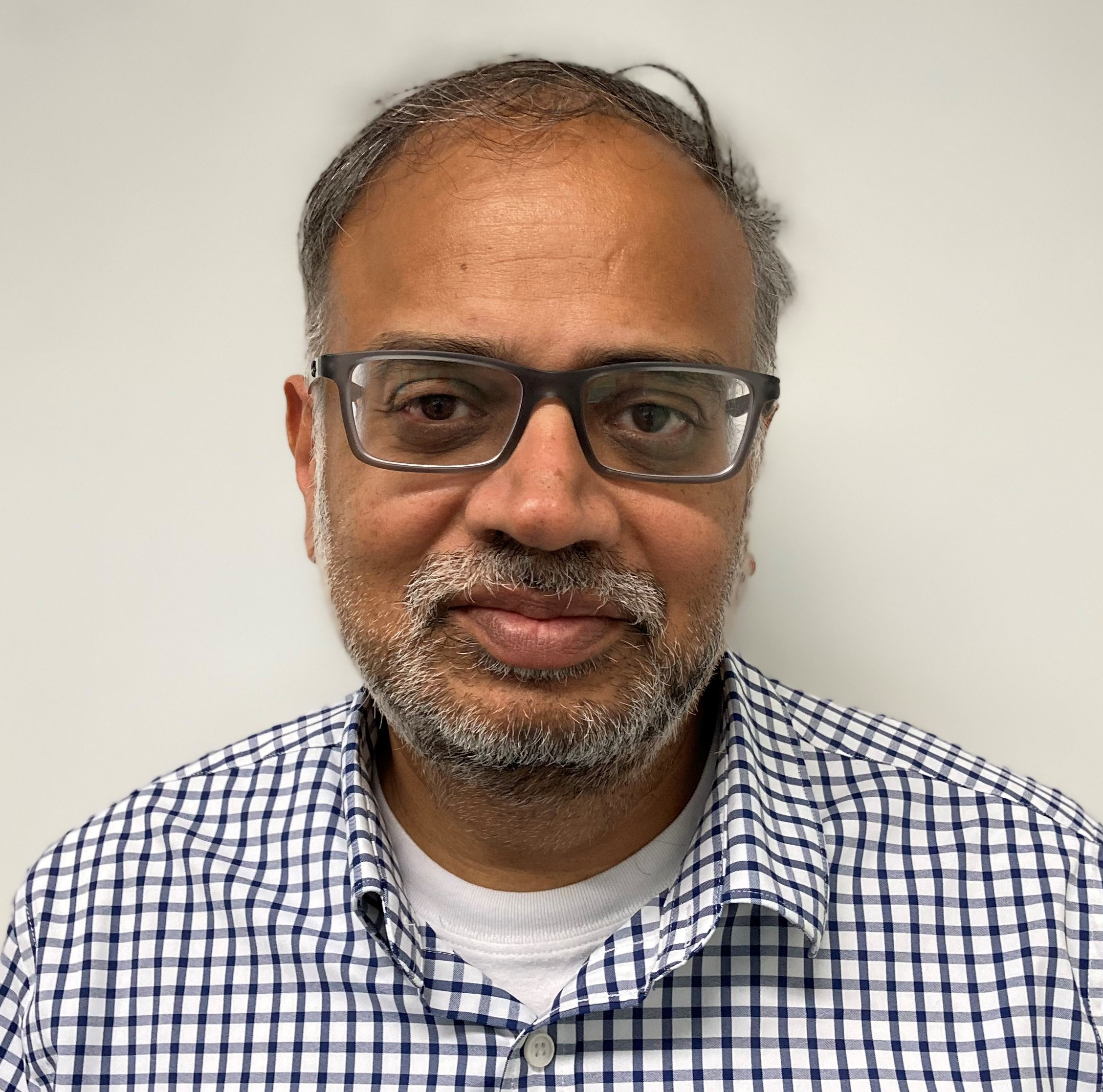}}]{Murti V. Salapaka}
(Fellow, IEEE) received the bachelor’s degree in mechanical engineering from the Indian Institute of Technology, Madras, in 1991, the master’s and Ph.D. degrees from the University of California at Santa Barbara, Santa Barbara, 1993 and 1997, respectively. He was with the Electrical Engineering Department, Iowa State University from 1997 to 2007. He is currently the Vincentine Hermes Luh Chair Professor with the Electrical and Computer Engineering Department, University of Minnesota, Minneapolis. He is a recipient of the NSF CAREER Award in 1998 and the ISU–Young Engineering Faculty Research Award in 2001.
\end{IEEEbiography}

\end{document}